\documentclass{article}%
\usepackage{amsmath}
\usepackage{amsfonts}
\usepackage{amssymb}
\usepackage{graphicx}%
\providecommand{\U}[1]{\protect\rule{.1in}{.1in}}

\begin{document}

\title{Localization and the interface between quantum mechanics, quantum field theory
and quantum gravity I\\(The two antagonistic localizations and their asymptotic compatibility)\\{\small dedicated to the memory of Rob Clifton}\\{\small Studies in History and Philosophy of Modern Physics 41 (2010)
104--127}}
\author{Bert Schroer\\CBPF, Rua Dr. Xavier Sigaud 150 \\22290-180 Rio de Janeiro, Brazil\\and Institut fuer Theoretische Physik der FU Berlin, Germany}
\maketitle
\tableofcontents

\begin{abstract}
It is shown that there are significant conceptual differences between QM and
QFT which make it difficult to view the latter as just a relativistic
extension of the principles of QM. At the root of this is a fundamental
distiction between Born-localization in QM (which in the relativistic context
changes its name to Newton-Wigner localization) and \textit{modular
localization} which is the localization underlying QFT, after one separates it
from its standard presentation in terms of field coordinates. The first comes
with a probability notion and projection operators, whereas the latter
describes causal propagation in QFT and leads to thermal aspects of locally
reduced finite energy states. The Born-Newton-Wigner localization in QFT is
only applicable asymptotically and the covariant correlation between
asymptotic in and out localization projectors is the basis of the existence of
an invariant scattering matrix.

In this first part of a two part essay the modular localization (the intrinsic
content of field localization) and its philosophical consequences take the
center stage. Important physical consequences of vacuum polarization will be
the main topic of part II. Both parts together form a rather comprehensive
presentation of known consequences of the two antagonistic localization
concepts, including the those of its misunderstandings in string theory.  

When the first version of this paper was submitted to hep-th it was
immediately removed and against its QFT content placed on phys. gen without
any possibility to cross list. 

\end{abstract}

\section{Introductory remarks}

Ever since the discovery of quantum mechanics (QM), the conceptual differences
between classical theory and QM have been the subject of fundamental
investigations with profound physical and philosophical consequences. But the
conceptual relation between quantum field theory (QFT) and QM, which is at
least as challenging and rich of surprises, has not received the same amount
of attention and scrutiny, and often the subsuming of QFT under "relativistic
QM" nourished prejudices and prevented a critical foundational debate. Apart
from some admirable work on the significant changes which the theory of
measurements must undergo in order to be consistent with the structure of QFT,
which emanated from people who are or have been affiliated with the Philosophy
of Science Department of the University of Pittsburgh \cite{Rob}%
\cite{Hal}\cite{Rue}, as well as some related deep mathematical and conceptual
work from quantum field theorists \cite{Werner}\cite{Sum1}\cite{Sum2}, this
subject has remained in the mind of a few individuals working on the
foundations of QFT and is still far from being part of the collective
knowledge of the foundation of QT community.

Often results of this kind which involve advanced knowledge of QFT do not
attract much attention even when they have bearings on the foundations of QT
as e.g. the issue of \textit{Bell states} in \textit{local quantum physics}
(LQP\footnote{We use this terminology instead of QFT if we want to direct the
reader's attention away from the textbook Lagrangian quantization towards the
underlying principles \cite{Haag}. QFT (the content of QFT textbooks) and LQP
deal with the same physical principles but LQP is less comitted to a
particular formalism (Lagrangian quantization, functional integrals) and
rather procures always the most adaequate mathematical concepts for their
implementation. It includes of course all the results of the standard
perturbative Lagrangian quantization but presents them in a conceptually and
mathematically more satisfactory way. Most of the subjects in this article are
outside of textbook QFT.}) \cite{Su} or the important relations between causal
disjointedness with the existence of uncorrelated states as well as the issue
to what extent causal independence is a consequence of statistical
independence \cite{Bu-Su}. The reason is not so much a lack of interest but
rather that QFT is often thought to be just a kind of relativistic quantum
mechanics and that possible differences are of a technical nature. This may
explain why there has been an amazing lack of balance between the very
detailed and sophisticated literature about interpretational aspects of QM and
its relation with quantum information theory (aiming sometimes at some very
fine, if not to say academic/metaphoric points e.g. the multiworld
interpretation), and the almost complete lack of profound interpretive
activities about our most fundamental quantum field theory of matter. Although
the name QT usually appears in the title of foundational papers, this mostly
hides the fact that they deal exclusively with concepts from QM leaving out QFT.

If on the other hand some foundational motivated quantum theorist become aware
of the deep conceptual differences between particles and fields, they tend to
look at them as antagonistic and create a battleground; the fact that they are
fully compatible where for physical reasons they must agree, namely in the
asymptotic region of scattering theory, remains often uncommented.

The aim of this essay is to show that at the root of these differences there
are two localization concepts: the quantum mechanical Born-Newton-Wigner
localization and the modular localization of LQP. The BNW localization is not
Poincar\'{e} covariant but attains this property in a certain asymptotic limit
namely the one on which scattering theory is founded. Modular localization on
the other hand is causal and covariant at all distances but provides no
projectors on subspaces as they arise from spectral decompositions of
selfadjoint or unitary operators, instead the linear spaces of localized
states are usually dense in the Hilbert space of all states. One of the aims
of this article is to collect some facts which, somewhat oversimplified, show
that besides sharing the notion of Hilbert space, operators, states and
Planck's constant $\hslash,$ QM and QFT are conceptually worlds
apart\footnote{For more illustrations of this point see the concluding remaks
in part II.} and yet they harmonize perfectly in the asymptotic region of
scattering theory.

For a long time the subtle distinction between the non-covariant BNW
localization (based on the existence of a position operator) and the
autonomous covariant localization concept of QFT was insufficiently
understood. It has been claimed (private communication by Rudolf Haag) that
the reason for Wigner, who together with Jordan significantly enriched QFT, to
become later disenchanted with this theory was that he failed to obtain a
covariant localization concept which was able to directly connect his
representation theory of the Poincar\'{e} group with QFT \footnote{This is
precisely what modular localization achieves (section 6).}. The application of
the non-covariant and hence frame-dependent BNW localization to finite
distances leads to incorrect results in particular to superluminal phenomena.
But only after the publication of an article \cite{Heger}, in which it was
claimed that Fermi's result about his two-atom Gedankenexperiment contradicts
understandings about spacetime localization and signal propagation, came the
issue of BNW- and modular- localization to a climax. The editor of
\textit{Nature} at that time wrote an article, in which among other aspects of
superluminal propagation, the issue of time machines was discussed.

After this there was no holding back; an avalanche of articles about
superluminal journeys and time machines entered the worldpress; the fact that
the original article appeared in a highly reputable journal, and its content
was reinforced and reprocessed for a scientifically interested public by a
well-known scientific journalist, made the topic irresistible to the general
public, especially since it had the right mixture of trustworthy origin with
sensational content.

Fortunately this was not the end of this episode. The same journal which
published the article based on the use of BNW localization accepted a second
article \cite{Bu-Yn} in which Fermi's conclusions about finite propagation
speed were reinforced on the basis of \textit{modular localization}%
\footnote{In the publications during the 90s, this terminology was not yet
available. It was sufficient to simply think in terms of the kind of
localization which is intrinsic to pointlike covariant fields.}. This episode
underlines the subtlety of localization in QFT and most of the content of both
parts of this essay will consist in explaining why this is such a delicate
problem which led to many misunderstandings.

It is not our intention to present a new axiomatic setting (for an older
presentation of the existing one see \cite{Haag}). Such a goal would be too
ambitious in view of the fact that we are confronting a theory where, in
contradistinction to QM, no conceptual closure is yet in sight. Although there
has been some remarkable nonperturbative progress concerning constructive
control (i.e. solving the existence problem) of models based on modular
theory, the main knowledge about models of QFT is still limited to numerically
successful, but nevertheless diverging perturbative series.

Here the more modest aim is to collect some either unknown or little known
facts which could present some food for thoughts about a more inclusive
measurement theory, including all of quantum theory (QT) end not just QM. On
the other hand one would like to improve the understanding about the interface
between QFT in CST curved spacetime (CST) and the still elusive quantum
gravity (QG).

Since both expressions QFT and LQP are used do denote the same theory, let me
emphasize again that there is no difference in the physical aims since LQP
originated from QFT and incorporated all concepts and computational results of
QFT including renormalized perturbation theory; LQP is used instead of QFT
whenever the conceptual level of the presentations gets beyond what the reader
is able to find in standard textbooks of QFT, more specifically whenever one
is interested in nonperturbative mathematically controlled constructions of
models in terms of intrinsic ("field-coordinatization independent")
structures. There is one recommendable exception, namely Rudolf Haag's book
"Local Quantum Physics" \cite{Haag}; but in a fast developing area of particle
physics two decades (referring to the time it was written) are a long time. \ 

The paper consists of two parts, the first is entirely dedicated to the
exposition of the differences between (relativistic\footnote{In order to show,
that by making QM relativistic, one does not remove the fundamental
differences with QFT, the next section will be on the relativistic setting of
"direct particle interactions".}) QM and LQP and their coexistence at large
time separations within the setting of scattering theory. The second part,
which will appear as a separate contribution, deals with thermal and entropic
consequences of vacuum polarization caused by causal localization, as well as
some consequences for QFT in CST. A QG theory does not yet exist, but a
profound understanding of those foundational aspects is expected to be
important to arrive at one.

The sections of the paper at hand are as follows. The first sections presents
the little known theory of \textit{direct particle interactions }(DPI), a
framework which incorporates all those \ properties of a relativistic theory
which one is able to formulate solely in terms of relativistic particles; some
of them already appeared in the pre Feynman S-matrix work of E.C.G.
St\"{u}ckelberg. In contradistinction to nonrelativistic QM where the cluster
factorization follows from the additivity of two-particle interactions, its
enforcement in DPI requires more refined arguments. As a closely related
result, DPI does not allow a second quantization presentation, even though it
is a perfect legitimate multiparticle theory in which n-particles are linked
to n+1 particles by cluster factorization. Most particle physicists tend to
believe that a relativistic particle theory, consistent with macro-causality
and a Poincar\'{e}-invariant S-matrix, must be equivalent to QFT\footnote{The
related folklore one finds in the literature amounts to the dictum:
relativistic quantum theory of particles + cluster factorization property =
QFT. Apparently this conjecture goes back to S. Weinberg.}, therefore it may
be helpful to show that this is not correct. DPI theories fulfill all the
physical requirements which one is able to formulate solely in terms
relativistic particles without recourse to fields, as Poincar\'{e} covariance,
unitary and macro-causality of the resulting S-matrix (which includes cluster factorization).

In this way one learns to appreciate the fundamental difference between
quantum theories which have no algebraically built-in maximal velocity and
those which have. As a quantum mechanical theory DPI only leads to statistical
"effective" finite velocity propagation for asymptotically large time-like
separations between localized events as they occur in scattering theory. With
other words the causal propagation between Born-localized events is a
macroscopic phenomenon for which, in analogy to the acoustic velocity in QM,
the large time behavior of dissipating wave packets is important, whereas in
QFT the maximal velocity is imprinted into the algebraic structure. DPI does
not possess covariant local operators, the only covariant object is the
Poincar\'{e} invariant S-matrix; from this viewpoint DPI is an S-matrix
theory. If one uses it outside of asymptotic propagation, one of course finds
the violation of causality which led to misunderstanding of Fermi's claim
\cite{Heger} which was subsequently corrected with the (implicit) use of
modular theory \cite{Bu-Yn}.

At the root of the QM-QFT (particle-field) antagonism is the existence of two
very different concepts of localization namely the \textit{Born localization}%
\footnote{It is interesting to note that Born introduced the probability
concept in QM in the context of the Born approximation of what we call
nowadays the cross section and not of the Schroedinger wave function
\cite{Born}. With other words he introduced it in the asymptotic region where
it is indidpensible and where the BNW localization becomes independent of the
reference frame.} (which is the only localization for QM), and the
\textit{modular localization} which underlies the causal locality in QFT. The
Born localization and the related position operator has been adapted to the
covariant normalization of relativistic wave functions in a paper by Newton
and Wigner \cite{N-W} (whereupon it becomes frame-dependent) and will
henceforth be referred to as the BNW localization. Whereas relativistic QM
permits only the BNW localization, QFT needs both, the modular
localization\footnote{Modular localization is the same as the causal
localization inherent in QFT after one liberates the letter from the
contingencies of particular selected fields. It is a property of the local
equivalence class of relatively loca; fields (the Borchers class) or of the
associated sysrem of local algebras. If one considers, as it is done in
algebraic QFT, the local fields as coordinatizations of the local algebras,
modular localization is independent of the "field coordinatization".} in order
to implement causal propagation and the BNW localization to get to the
indispensable asymptotic scattering probabilities (cross sections). Without
the BNW localization QFT would remain a beautiful mathematical construct with
no accessible physical content; on the other hand without modular localization
QFT would not have interaction-induced vacuum polarization and its description
of reality at finite distances would contain acausal poltergeist-daemons of
the kind mentioned above. Note that we avoid the phrase "particle-wave
dualism" because in our understanding this issue has been solved in the
transformation theory showing that Schroedinger's wave function formalism is
equivalent to the algebraic formulation in terms of p.q \ operators or in the
relativistic context that Wigner's particle positive energy representation
approach to particles is uniquely functorially related to free fields or the
related spacetime localized sytem of algebras (sections 6,7). The
particle-field problem starts only in the presence of interacting QFTs. 

Particles are objects with a well-defined ontological status, whereas (basic
and composite) fields form an infinite set of coordinatizations which generate
the local algebras. By this we mean that particles are the truly real and
unique objects which are subject to direct observations and independent of any
"field-coordinatization", a property which is not derogated by the fact that
their existence is only an asymptotic one. What is referred to as an
(asymptotic) n-particle state is a state in which n well separated coincidence
counters (in a world cobbled with counters) click simultaneously and apart
from the localization of the counters, the number n does not change at later
times and the particle cross section is frame independent.

Quantum fields on the other hand have a more fleeting and less individual
existence and there are always infinitely many fields which are associated
with one particle. This finds its expression in the terminology "interpolating
fields" used in the LSZ scattering theory of the 50s. But as epistemic
entities fields or local algebras are indispensable since all our intuition
about local interactions and their causal localization properties is injected
on the level of fields or directly into the local observable algebras which
they generate. Without Born localization and the associated projectors, there
would be no scattering theory leading to cross sections and hence QFT would be
reduced to just a physically anemic mathematical playground.

In contradistinction to DPI, in interacting QFT there is no way in which in
the presence of interactions the notion of \textit{particles at finite times}
can be saved. The statement that an isolated relativistic particle cannot be
localized below its Compton wave length refers to the (Newton-Wigner
adaptation of the) Born localization and, as all statements involving Born
localization, it is meant in an \textit{effective} probabilistic sense. Only
in the timelike asymptotic limit between two events the BNW localization
becomes a sharp geometric relations in terms of momenta with c being the
maximal velocity which is independent of the reference frame; fortunately this
is precisely what one needs to obtain a Poincar\'{e} invariant macrocausal S-matrix.

The maximal velocity in the sense of asymptotic expectations in suitable
states of relativistic particle theories plays a similar role as acoustic
velocity in nonrelativistic QM which leads to (material-dependent) acoustic
velocities. Placing our interpretation in the context of prior work on this
subject \cite{Hal}\cite{Fleming}, our conclusion is that neither
"Reeh-Schlieder defeats Newton-Wigner", nor does Newton-Wigner "meet"
Reeh-Schlieder in the nonaymptotic spacetime region, rather \textit{both
indispensable localization schemes} approximate each other asymptotically at
$t\rightarrow\pm\infty$ where the Newton-Wigner localization becomes
covariant, the results of macro-causality coalesce with those of
micro-causality and the modular localization shares the asymptotic probability
notion with BNW, i.e. no defeat of either one, but harmony at the only place
where they can meet and remain faithful to their principles.

The next section contains some remarks about the history of the growing
awareness about properties which separate QM from QFT. This is followed in
section 3 with the presentation of a little known consistent setting of
interacting relativistic particles without fields: the direct particle
interaction theory (DPI) by Coester and Polyzou. Sections 4 and 5 focus on the
radical difference between the Newton-Wigner (NW) localization (the name for
the Born localization after the adaptation to the relativistic particle
setting) and the localization which is inherent in QFT, which in its intrinsic
form, i.e. liberated from singular pointlike "field coordinatizations", is
referred to as \textit{modular localization }\cite{BGL}\cite{Sch}%
\cite{MSY}\textit{. }The terminology\textit{\ }has its origin in the fact that
it is backed up by a mathematical theory within the setting of operator
algebras which bears the name Tomita-Takesaki\footnote{Tomita was a Japanese
mathematician who discovered the main properties of the theory in the first
half of the 60s, but it needed a lot of polishing by Takesaki in order to be
accepted.} \textit{modular theory.} Within the \ setting of thermal QFT,
physicists independently discovered various aspects of this theory
\cite{Haag}. Its relevance for causal localization was only spotted a decade
later \cite{Bi-Wi} and the appreciation of its role in problems of thermal
behavior at causal- and event- horizons and black hole physics had to wait
another decade \cite{Sew}.

Sections 6-10 are all centered around an in-depth exposition of various
aspects of modular localization, starting from the modular localization of
states and passing to its more restrictive algebraic counterpart. Among its
very recent application is the notion of semiinfinite spacelike string
localization which on the one hand settled the age old problem of the
appropriate localization for behind the Wigner infinite spin representation
but also shows that the object of string theory is really an infinite
component field (section 7).

The penultimate section presents LQP as the result of \textit{relative
positioning} of a finite (and rather small) number of \textit{monads} within a
Hilbert space; here we are using a terminology which Leibniz introduced in a
philosophical context but which makes a perfect match with the conceptual
structure of LQP once one goes beyond the classical quantization parallelism.
On the other hand this underlines the enormous conceptual distance between to
QM for which such concepts are not available. Whereas a single monad also
appears in different contexts, e.g. KMS states on open quantum systems (even
in QM) and the information theoretical interpretation of bipartite spin
algebras in suitable singular states \cite{Werner}\cite{Keyl-M}, the modular
positioning of several copies is totally characteristic for LQP. Although its
physical and mathematical content is quite different from Mermin's
\cite{Mermin} new look (the "Ithaca-interpretation" of QM) at quantum
mechanical reality exclusively in terms of correlations between subsystems,
the two concepts share the aspect of viewing reality in relational
terms\footnote{Mermin's relational idea remains however somewhat vague
compared to the mathematically very precise and physically complete
characterozation of QFT models via modular positioning.}. Mathematically a
monad in the sense of this article is the unique hyperfinite type III$_{1}$
factor algebra to which all local algebras in LQP are isomorphic, so all
concrete monads are copies of the abstract monad. Naturally a monade in
isolation is an abstract entity without structure, the reality emerges from
relations between monads within the same Hilbert space.

Whereas for Newton physical reality consisted of matter moving in a fixed
space according to a universal time, reality for Leibniz emerges from
interrelations between monads with spacetime serving as an ordering device.
The modular positioning of monads goes one step further in that even the
Minkowski spacetime together with its invariance group (the Poincar\'{e}
group) appears as a consequence of positioning in a more abstract sense namely
of a finite number of monads in a joint Hilbert space (subsection 7)
\cite{K-W}. For actual constructions of interacting LQP models it is however
advantageous to start with one monad and the action of the Poincar\'{e} group
on it \cite{Lech1}\cite{Lech2}.

The algebraic structure of QM on the other hand, relativistic or not, has no
such monad structure; the global algebra as well as all Born-localized
subalgebras in ground states are always of type I i.e. either the algebra of
all bounded operators $B(H)$ in an appropriate Hilbert space or multiples
thereof. Correlations are characteristic features of quantum mechanical
states, whereas for the characterization of a QM system global operators as
the Hamiltonian are indispensable.

Part I of this essay closes with a section on the split inclusion which shows
how in the ubiquitous presence of vacuum polarization some of the notion known
from QM (tensor factorization of disjoint subsystem, entanglement) can be
recuperated. The second part will present many more applications of modular
localization and the split property notably those related to thermal and
entropic properties which are of potential astrophysical and cosmological relevance.

\section{ Historical remarks on the interface between QM and QFT}

Shortly after the discovery of field quantization in 1925 \cite{Du-Jan}%
\cite{S-Jo}, there were two opposed viewpoints about its content and purpose
of relativistic QT represented by Dirac and Jordan \cite{Dar}. Dirac
maintained that quantum theory should stand for \textit{quantizing a classical
reality} which meant field quantization for electromagnetism and quantization
of classical mechanics for particles. Jordan, on the other hand, proposed an
uncompromising field quantization point of view; his guiding theme was that
all what can be quantized should be quantized, independent of whether there is
a classical reality or not\footnote{His radicality about subjecting everything
which mathematically can be quantized (including De Broglie matter waves) to
the new quantum recipe was also met with in certain cases justified
criticism.especially when he together with Klein second quantized the
Schroedinger equation (why quantize something again which was already
"quantum").}. The more radical field quantization including particles finally
won the argument, but ironically it was Dirac's particle setting (the hole
theory) and not Jordan's version of "Murphy's law" ("everything which can be
quantized must be quantized") to all field objects which contributed the
richest structural property to QFT, namely charge-anticharge symmetry leading
to the necessary presence of antiparticles.

It was also Dirac's hole theory setting in which the first perturbative QED
computations (which entered the textbooks of Heitler and Wenzel) were done,
before it was recognized that this setting was not really consistent. This
inconsistency \ showed up in problems involving renormalization in which
\textit{vacuum polarization} plays the essential role. The successful
perturbative renormalization of QED in the charge symmetric description was
also the end of hole theory as well as the start of Dirac's late acceptance of
QFT as the general setting for relativistic particle physics (at the beginning
of the 50s). This shows that before serious errors misled particle physics
into the present crisis \cite{foun} there were fascinating errors whose
profound understanding led too a deep enrichment and belong to our precious heritage.

Vacuum polarization is a very peculiar phenomenon which in the special context
of currents and the associated local charges of a complex free Bose field was
noticed already in the 30s by Heisenberg \cite{Hei}. But only when Furry and
Oppenheimer \cite{Fu-Op} studied perturbative interactions of Lagrangian
fields and became aware to their amazement that the Lagrangian field applied
to the vacuum created inevitably particle-antiparticle pairs in addition to
the expected one-particle state, the subtlety of the particle-field relation
within interacting QFT begun to be noticed. The number of these pairs increase
with the perturbative order, pointing towards the fact that in case of sharp
localization ("banging" with sharply localized operators onto the vacuum) one
has to deal with infinite polarization clouds containing arbitrary high energy
components. Since there is no position operator in QFT, there is a fortiori no
Heisenberg uncertainty relation. As a QFT substitute one may consider the
unbounded increase of localization entropy $Ent(\mathcal{O},\varepsilon)$
where $\mathcal{O}$ is the spacetime localization region and $\varepsilon$ the
split distance which creates a "fuzzy" surface. When $\varepsilon\rightarrow0$
the entropy diverges as $\frac{A(H(\mathcal{O}))}{\varepsilon^{2}}%
ln\sqrt{\frac{A(H(\mathcal{O}))}{\varepsilon^{2}}}$ where $A(H(\mathcal{O}))$
is the area of the causal horizon of $\mathcal{O},$ i.e. the sharper the
localization the bigger the localization entropy \cite{interfaceII}\cite{BMS}.

Whenever one tries in an interacting theory to create particles via local
disturbances of the vacuum, the vacuum polarization clouds corrupt the
observation of those particles which one intends to create, but after a
sufficient amount of time the particle content separates from the polarization
cloud. In the presence of interactions the notion of particles in local
regions at a fixed time is, strictly speaking, meaningless because even the
field with the "mildest" vacuum polarization taken from the class of all
possible relative local fields which all interpolate the same particle still
generates an infinite vacuum polarization cloud which sticks inseparably to
the particle of interest\footnote{Only if one allows noncompact localization
regions one is able to find "PFGs" i.e. operators which applied to the vacuum
generate one-particle states without polarization admixture. Wedge regions in
Minkowski space lead to the best compromise between particles and fields and
play a fundamental role in recent model constructions \cite{S3}\cite{BBS}%
\cite{Lech1} and is at the root of the crossing property \cite{foun}. For a
philosophical viewpoint see \cite{Fraser}.}. It is somewhat ironic that
particles, which are the main bridge between QFT and its laboratory reality
(and which are the basic objects of QM), have only an asymptotic existence as
incoming and outgoing asymptotic particle configurations.

In the next subsection it will be shown that relativistic QM in the form of
DPI, in contradistinction of what most particle theorists believe, can be
consistently formulated \cite{C-P} and this setting can even be extended to
incorporate creation and annihilation channels \cite{P}. This goes along way
to vindicate Dirac's relativistic particle viewpoint. But it does not
vindicate it completely, since theories which start as particle theories but
then lead to vacuum polarization as Dirac's hole theory are at the end
inconsistent unless one converts their content into a charge symmetric field
theoretic setting (in which case the connection with Dirac's whole theory is lost).

By contrasting QFT with DPI, one obtains a better appreciation of the
conceptual depth of QFT, in particular one becomes aware of its still
unexplored regions. DPI is basically a relativistic \textit{particle} setting
i.e. it deals only with properties which can be formulated in terms of
particles; this limits causality properties to \textit{macro-causality} i.e.
spacelike cluster factorization and timelike causal rescattering. Apart from
the fact that the multi-particle representation theory of the Poincar\'{e}
group is incompatible with the additivity of interaction terms which
complicates the implementation of the cluster factorization property and
prevents an elegant second quantization description in Wigner-Fock space, the
DPI setting is as well understood as nonrelativistic QM. In contrast nobody
who has studied QFT beyond a textbook level would claim that QFT is anywhere
near its closure. The last section illustrates this point by an unexpected new
abstract characterization of QFT which is different from any previous
axiomatic attempt.

\section{Direct particle interactions, relativistic QM}

The Coester-Polyzou \ theory of \textit{direct particle interactions} (DPI)
(where "direct" means "not field-mediated") is a relativistic setting in the
sense of representation theory of the Poincar\'{e} group which, among other
things, leads to a Poincar\'{e} invariant S-matrix. Every property which can
be formulated in terms of particles, as the cluster factorization into systems
with a lesser number of particles and other timelike aspects of
macrocausality, can be implemented in this setting. The S-matrix does however
not fulfill such analyticity properties as the crossing \cite{foun} property
whose derivation relies on the existence of local interpolating fields.

In contradistinction to the more fundamental locally covariant QFT, DPI is
primarily a phenomenological setting, but one which is consistent with every
property which can be expressed in terms of relativistic particles only. So
instead of approximating nonperturbative QFT in a metaphoric way outside
conceptional-mathematical control, the idea of DPI is to arrange
phenomenological calculations in such a way that at least the principles of
relativistic mechanics and macro-causality are maintained \cite{C-P}.

For the interaction of two relativistic particles the introduction of
relativistic interactions amounted to add to the free mass operator (the
Hamiltonian in the c.m. system) an interact which depends on the relative
position and momentum. The exigencies of representation theory of the
Poincar\'{e} group are then fulfilled and the cluster property stating that
$S\rightarrow\mathbf{1}$ for large spatial separation is a consequence of the
short ranged interaction. Assuming for simplicity identical scalar Bosons, the
c.m. invariant energy operator is $2\sqrt{p^{2}+m^{2}}~$and the interaction is
introduced by adding an interaction term $v$%

\begin{equation}
M=2\sqrt{\vec{p}^{2}+m^{2}}+v,~~H=\sqrt{\vec{P}^{2}+M^{2}}%
\end{equation}
where the invariant potential $v$ depends on the relative c.m. variables $p,q$
in an invariant manner i.e. such that $M$ commutes with the Poincar\'{e}
generators of the 2-particle system which is a tensor product of two
one-particle systems.

One may follow Bakamjian and Thomas (BT) \cite{BT} and choose the Poincar\'{e}
generators in their way so that the interaction only appears in the
Hamiltonian. Denoting the interaction-free generators by a subscript $0,$ one
arrives at the following system of two-particle generators%
\begin{align}
\vec{K}  &  =\frac{1}{2}(\vec{X}_{0}H+H\vec{X}_{0})-\vec{J}\times\vec{P}%
_{0}(M+H)^{-1}\\
\vec{J}  &  =\vec{J}_{0}-\vec{X}_{0}\times\vec{P}_{0}\nonumber
\end{align}

The interaction $v$ may be taken as a \textit{local} function in the relative
coordinate which is conjugate to the relative momentum $p$ in the c.m. system;
but since the scheme anyhow does not lead to local differential equations,
there is not much to be gained from such a choice. The Wigner canonical spin
$\vec{J}_{0}$ commutes with $\vec{P}=\vec{P}_{0}$ and $\vec{X}=\vec{X}_{.0}$
and is related to the Pauli-Lubanski vector $W_{\mu}=\varepsilon_{\mu\nu
\kappa\lambda}P^{\nu}M^{\kappa\lambda}$ .

As in the nonrelativistic setting, short ranged interactions $v$ lead to
M\o ller operators and S-matrices via a converging sequence of unitaries
formed from the free and interacting Hamiltonian%
\begin{align}
\Omega_{\pm}(H,H_{0})  &  =\lim_{t\rightarrow\pm\infty}e^{iHt}e^{-H_{0}t}\\
\Omega_{\pm}(M,M_{0})  &  =\Omega_{\pm}(H,H_{0})\label{sec}\\
S  &  =\Omega_{+}^{\ast}\Omega_{-}\nonumber
\end{align}
The identity in the second line is the consequence of a theorem which say that
the limit is not affected if instead of $M$ one takes take a positive function
of $M$ (\ref{sec}) as $H(M),$ as long as $H_{0}$ is the same function of
$M_{0}.$ This insures the the asymptotic \textit{frame-independence of objects
as the M\o ller operators and the S-matrix }but not necessarily that of semi
asymptotic operators as formfactors of local operators between ket in and bra
out particle states. Apart from this \textit{identity for operators and their
positive functions} (\ref{sec}) which is not needed in the nonrelativistic
scattering, the rest behaves just as in nonrelativistic scattering theory. As
in standard QM, the 2-particle cluster property is the statement that
$\Omega_{\pm}^{(2)}\rightarrow\mathbf{1,}$ $S^{(2)}\rightarrow\mathbf{1,}$
i.e. the scattering formalism is identical. In particular the two particle
cluster property, which says that for short range interactions the S-matrix
approaches the identity if one separates the center of the wave packets of the
two incoming particles, holds also for the relativistic case.

The implementation of clustering is more delicate for three particles as can
be seen from the fact that the first attempts were started in 1965 by Coester
\cite{Coe} and considerably later generalized (in collaboration with Polyzou
\cite{C-P}) to an arbitrary high particle number. To anticipate the result
below, DPI leads to a consistent scheme which fulfills cluster factorization
but it has no useful second quantized formulation so it may stand accused of
lack of elegance; one is inclined to view less elegant theories also as less
fundamental. It is also more nonlocal and nonlinear than Galilei-incariant QM,
This had to be expected since adding interacting particles does not mean
adding up interactions as in Schroedinger QM.

The BT form for the generators can be achieved inductively for an arbitrary
number of particles. As will be seen, the advantage of this form is that in
passing from n-1 to n-particles the interactions add after appropriate
Poincar\'{e} transformations to the joint c.m. system and in this way one ends
up with Poincar\'{e} group generators for an interacting n-particle system.
But for $n>2$ the aforementioned subtle problem with the cluster property
arises; whereas this iterative construction in the nonrelativistic setting
complies with cluster separability, this is not the case in the relativistic context.

This problem shows up for the first time in the presence of 3 particles
\cite{Coe}. The BT iteration from 2 to 3 particles gives the 3-particle mass operator%

\begin{align}
M  &  =M_{0}+V_{12}+V_{13}+V_{23}+V_{123}\label{add}\\
V_{12}  &  =M(12,3)-M_{0}(12;3),~M(12,3)=\sqrt{\vec{p}_{12,3}^{2}+M_{12}^{2}%
}+\sqrt{\vec{p}_{12,3}^{2}+m^{2}}\nonumber
\end{align}
and the $M(ij,k)$ result from cyclic permutations. Here $M(12,3)$ denotes the
3-particle invariant mass in case the third particle is a \textquotedblleft
spectator\textquotedblright,\ which by definition does not interact with 1 and
2. The momentum in the last line is the relative momentum between the
$(12)$-cluster and particle $3$ in the joint c.m. system and $M_{12}$ is the
associated two-particle mass i.e. the invariant energy in the $(12)$ c.m.
system. Written in terms of the original two-particle interaction $v,$ the
3-particle mass term appears nonlinear.

As in the nonrelativistic case, one can always add a totally connected
contribution. Setting this contribution to zero, the 3-particle mass operator
only depends on the two-particle interaction $v.~$But contrary to the
nonrelativistic case, the BT generators constructed with $M$ as it stands does
not fulfill the cluster separability requirement. The latter demands that if
the interaction between two clusters is removed, the unitary representation
factorizes into that of the product of the two clusters.

One expects that shifting the third particle to infinity will render it a
spectator and result in a factorization $U_{12,3}\rightarrow U_{12}\otimes
U_{3}$. Unfortunately what really happens is that the $(12)$ interaction also
gets switched off i.e. $U_{123}\rightarrow U_{1}\otimes U_{2}\otimes U_{3}$ .
The reason for this violation of the cluster separability property, as a
simple calculation (using the transformation formula from c.m. variables to
the original $p_{i}$, i = 1, 2, 3) shows \cite{C-P}), is that although the
spatial translation in the original system (instead of the $12,3$ c.m. system)
does remove the third particle to infinity as it should, unfortunately it also
drives the two-particle mass operator (with which it does not commute) towards
its free value which violates clustering.

In other words the BT produces a Poincar\'{e} covariant 3-particle interaction
which is additive in the respective c.m. interaction terms (\ref{add}), but
the Poincar\'{e} representation $U$ of the resulting system will not be
cluster-separable. However this is the time for intervention of a saving
grace: \textit{scattering equivalence}.

As shown first in \cite{Coe}, even though the 3-particle representation of the
Poincar\'{e} group arrived at by the above arguments violates clustering, the
3-particle S-matrix computed in the additive BT scheme turns out to have the
cluster factorization property. But without implementing the correct cluster
factorization also for the 3-particle Poincar\'{e} generators there is no
chance to proceed to a clustering 4-particle S-matrix.

Fortunately there always exist unitaries which transform BT systems into
cluster-separable systems \textit{without affecting the S-matrix}. Such
transformations are called \textit{scattering equivalences. }They were first
introduced into QM by Sokolov \cite{So} and their intuitive content is related
to a certain insensitivity of the scattering operator under quasilocal changes
of the quantum mechanical description at finite times. This is reminiscent of
the insensitivity of the S-matrix against local changes in the interpolating
field-coordinatizations\footnote{In field theoretic terminology this means
changing the pointlike field by passing to another (composite) field in the
same equivalence class (Borchers class) or in the setting of AQFT by picking
another operator from a local operator algebra.} in QFT by e.g. using
composites instead of the Lagrangian field.

The notion of scattering equivalences is conveniently described in terms of a
subalgebra of \textit{asymptotically constant operators} $C$ defined by%
\begin{align}
\lim_{t\rightarrow\pm\infty}C^{\#}e^{iH_{0}t}\psi &  =0\\
\lim_{t\rightarrow\pm\infty}\left(  V^{\#}-1\right)  e^{iH_{0}t}\psi &
=0\nonumber
\end{align}
where $C^{\#}$ stands for both $C$ and $C^{\ast}$. These operators, which
vanish on dissipating free wave packets in configuration space, form a
*-subalgebra which extends naturally to a $C^{\ast}$-algebra $\mathcal{C}$. A
scattering equivalence is a unitary member $V\in$ $\mathcal{C}$ which is
asymptotically equal to the identity (the content of the second line).
Applying this asymptotic equivalence relation to the M\o ller operator one obtains%

\begin{equation}
\Omega_{\pm}(VHV^{\ast},VH_{0}V^{\ast})=V\Omega_{\pm}(H,H_{0})
\end{equation}
so that the $V$ cancels out in the S-matrix. Scattering equivalences do
however change the interacting representations of the Poincar\'{e} group
according to $U(\Lambda,a)\rightarrow VU(\Lambda,a)V^{\ast}.$

The upshot is that there exists a clustering Hamiltonian $H_{clu}$ which is
unitarily related to the BT Hamiltonian $H_{BT}$ i.e. $H_{clu}=BH_{BT}B^{\ast
}$ such that $B\in\mathcal{C}~$is uniquely determined in terms of the
scattering data computed from $H_{BT}.$ It is precisely this clustering of
$H_{clu}$ which is needed for obtaining a clustering 4-particle S-matrix which
is cluster-associated with the $S^{(3)}$. With the help of $M_{clu}$ one
defines a 4-particle interaction following the additive BT prescription; the
subsequent scattering formalism leads to a clustering 4-particle S-matrix and
again one would not be able to go to n=5 without passing from the BT to the
cluster-factorizing 4-particle Poincar\'{e} group representation. Coester and
Polyzou showed \cite{C-P} that this procedure can be iterated and doing this
one arrives at the following statement

\textbf{Statement}: \textit{The freedom of choosing scattering equivalences
can be used to convert the Bakamijan-Thomas presentation of multi-particle
Poincar\'{e} generators into a cluster-factorizing representation. In this way
a cluster-factorizing S-matrix }$S^{(n)}$\textit{ associated to a BT
representation }$H_{BT}$\textit{ (in which clustering mass operator }%
$M_{clu}^{(n-1)}$\textit{ was used) leads via the construction of }%
$M_{clu}^{(n)}$\textit{ to a S-matrix }$S^{(n+1)}$\textit{ which clusters in
terms of all the previously determined }$S^{(k)},k<n.$ \textit{The use of
scattering equivalences prevents the existence of a 2}$^{nd}$\textit{
quantized formalism. }

For a proof we refer to the original papers \cite{C-P}\cite{P}. In passing we
mention that the minimal extension, (i.e. the one determined uniquely in terms
of the two-particle interaction $v$) from n to n+1 for $n>3,$ contains
\textit{connected 3-and higher particle interactions} which are nonlinear
expressions (involving nested roots) in terms of the original two-particle
$v.~$This is another unexpected phenomenon as compared to the nonrelativistic case.

This theorem shows that it is possible to construct a relativistic theory
which only uses particle concepts, thus correcting an old folklore which says
relativity + clustering = QFT. Whether one should call this DPI theory
"relativistic QM" or just a relativistic S-matrix theory in a QM setting is a
matter of taste; it depends on what significance one attributes to those
unusual scattering equivalences. In any case it defines a \textit{relativistic
S-matrix setting }with the correct particle behavior i\textit{.}e\textit{.
}all properties which one is able to formulate in terms of particles (without
the use of fields) as unitarity, Poincar\'{e} invariance and macrocausality
are fulfilled. In this context one should also mention that the S-matrix
bootstrap approach never addressed these macro-causality problems of the DPI
approach; it was a grand self-deluding design for a unique theory of all
non-gravitational interactions in which important physical details were
arrogantly ignored.

As mentioned above Coester and Polyzou also showed that this relativistic
setting can be extended to processes which maintain \textit{cluster
factorization in the presence of a finite number of creation/annihilation
channels}, thus demonstrating, as mentioned before, that \textit{the mere
presence of particle creation is not characteristic for QFT} but rather the
presence of infinite vacuum polarization clouds from "banging" with localized
operators onto the vacuum (see section 7). Different from the nonrelativistic
Schroedinger QM, the superselection rule for masses of particles which results
from Galilei invariance for nonrelativistic QM does not carry over to the
relativistic setting; in this respect DPI is less restrictive than its
Galilei-invariant QM counterpart where such creation processes are forbidden.

One may consider the DPI setting of Coester and Polyzou as that scheme which
results from implementing the mentioned particle properties within a
n-particle Wigner representation setting in the presence of interaction
\cite{C-P}, it is the only relativistic QM which is consistent with
macrocausality\footnote{Macrocausality consists of spacelike clustering and
timelike causal rescattering (St\"{u}ckelberg); it is the only causality which
one is able to fulfill in terms of particles only (without fields).}.
Apparently the work of these mathematical nuclear physicists has not been
noticed by particle physicists probably since the authors have published most
of their results in nuclear physics journals. What makes it worthwhile to
mention this work is that even physicists of great renown as Steven Weinberg
did not believe that such a theory exists because otherwise they would not
have conjectured that the implementation of cluster factorization properties
in a relativistic setting leads to QFT \cite{Wein}.

Certain properties which are consequences of locality in QFT and can be
formulated but not derived in a particle setting as the TCP symmetry, the
spin-statistics connection and the existence of anti-particles, can be added
"by hand" to the DPI setting. Other properties which are on-shell relics of
locality which QFT imprints on the S-matrix and which require the notion of
analytic continuation in particle momenta (as e.g. the crossing property for
formfactors) cannot be implemented in the QM setting of DPI.

\section{First brush with the intricacies of the particles-field problems in
QFT}

In contrast to QM (Schr\"{o}dinger-QM or relativistic DPI), interacting QFT
does not admit a particle interpretation at finite times\footnote{Although the
one-particle states and their multiparticle counterparts are global states in
the Hilbert space, they are not accessible by acting locally on the vacuum.
Scattering theory is the only known nonlocal intervention.}. If it would not
be for the asymptotic scattering interpretation in terms of incoming/outgoing
particles associated with the free in/out fields, there would be hardly
anything of a non-fleeting measurable nature. In QFT in CST and thermal QFT
where even this asymptotically valid particle concept is missing, the set of
conceivable measurements is essentially reduced to energy- and entropy-
densities in thermal states and in black hole states with event horizons as
well as of cosmological states describing the microwave background radiation.

Since the notion of particle is often used in a more general sense than in
this paper, it may be helpful to have a brief interlude on this issue. By
particle I mean an asymptotically stable object which leads through its
n-particle  tensor product structure to an asymptotically complete description
of the Hilbert space of a Minkowski spacetime QFT. It is precisely this
concept which furnishes QFT with a (LSZ, Haag-Ruelle) complete asymptotic
particle interpretation\footnote{The asymptotic completeness property was for
the first time established (together with a recent existence proof) in a
family of factorizing two-dimensional models (see the section on modular
localization) with nontrivial scattering.}, so that the Hilbert space of such
an interacting theory has a Fock space tensor structure. The physics behind is
the idea \cite{Ha-Sw} that if we were to "cobble" the asymptotic spacetime
region with counters which monitor coincidences/anticoincidences of
localization events (local deviations from the vacuum) after a collision of
two incoming particles has taken place, then the defining property of an
outgoing n-particle state is the stable n-fold coincidence/anticoincidence
(the latter in order to insure that we registered all particles) between n
counters. The intuitive idea is that after some time the n would not change
and the n-fold local excitations from the vacuum would move along trajectories
of free relativistic particles. would eventually remain stable because the far
removed localization centers would have ceased to interact and from there on
move freely. The occurrence rate of these coincidences as well as their
correlation with that of the incoming coincidences is independent of the frame
of reference even though BNW localization at finite spacetime regions is frame
dependent. In popular textbooks this is expressed as. the BNW localization
becomes "effectively" covariant for distances beyond a Compton wavelength
(exactly covariant only in the large time limit.) The Newton-Wigner adaptation
of the Born position operator would lead to genuinely Poincar\'{e} invariant
frame-independent transition probabilities between incoming and outgoing
Newton-Wigner localization events.

The particle concept in QFT is therefore precisely applicable where it is
needed, namely for asymptotically separated BNW-localized events for which the
probability interpretation and covariance become compatible. In fact the use
of the BNW localization for finite distances is known to lead to trouble in
form of unphysical superluminal effects; in that case one should formulate the
problem in the setting of the modular localization which has instead of
probabilities and projectors dense subsets of states (the Reeh-Schlieder
property \cite{Haag}).

Tying the particle concept in QFT to asymptotically stable coincidences of
counters can be traced back to a seminal paper by Haag and Swieca
\cite{Ha-Sw}. These authors noticed for the first time that the phase space
degree of freedom density in QFT, unlike that in QM, is not finite, rather its
cardinality is mildly infinite (the phase space is \textit{nuclear}). The
larger number of degrees of freedom in form of an enhanced phase space density
is yet another line of unexpected different consequences \cite{Swieca}%
\cite{Dybal2} resulting from the different localization concepts in QM and
QFT, but this interesting topic will not be pursued here.

Not all particles comply with this definition; in fact all electrically
charged particles are \textit{infraparticles} i.e. objects which are
asymptotically stable but in contrast to Wigner particles they are inexorably
attached to an unobserved cloud of infinitely many infrared photons leave a
mark of their presence even in the low energy part of the inclusive QED cross
section for charged infraparticles. The existence of an electron as a Wigner
particle associated with a sharp mass hyperboloid on top of a photon
background is a fiction which is incompatible with QED. Rather electrically
charged particles have instead of a mass shell delta function in their
Kallen-Lehmann two%
\'{}%
point function a cut which starts at $p^{2}=m^{2}~$which makes a precise
description of such \textit{infraparticles }\cite{infra} and their scattering
theory more involved. There is a difference between the direct application to
a theory for which the pole structure has been replaced by the infraparticle
cuts and the perturbative calculation which proceeds as if the mass shell
restriction makes sense (using ad hoc infrared cutoffs to combat
divergencies). Since the positivity of the K\"{a}llen-Lehmann measure forces
the cut singularity to be milder than the mass shell delta function, the LSZ
limits of the charged fields vanish. On the other hand the result of the
calculation of the inclusive infraparticle cross sections in the Boch-Nordsiek
model and the perturbative summation of leading infrared singularities in
\cite{YFS} lead to nonvanishing results only if the photon resolution is kept
finite; in particular one obtains a vanishing cross section for a finite
number of photons which is consistent with the trivial LSZ limits. Such
successful recipes hide the fact that the root of the problem is a radical
change of the particle concept which entails a fundamental adjustment
\cite{Porr}. These asymptotic attempt in momentum space hide the fact that
gauge theories are only useful for pointlike generated fields (field strength,
current) but not for the even more important charge fields which turn out to
be semiinfinite string-like generated. 

In contrast to Wigner particles which are representation theoretical objects
of the Poincar\'{e} group, infraparticles exist only in QED-like interacting
theories in which the charge obeys a quantum Gauss law holds. The most
dramatic differences between infraparticles and Wigner particles show up in
localization aspects. Whereas Wigner particles "are pointlike" i.e. have
pointlike generating wave functions, the sharpest localized generators for
infraparticles are semiinfinite stringlike. On a formal level this has been
known for a long time as expressed in the Dirac-Jordan-Mandelstam formulas in
which a Dirac spinor is multiplied by an exponential semiinfinite line
integral over the vectorpotential (\ref{DJM}). Their modern exposition would
be an important part of an essay about various localization concepts. However
the description of string-localized infraparticles is too subtle and would
require a presentation which goes much beyond the content of this essay. We
hope to return to issue in a separate paper.

It is the \textit{asymptotic particle structure} which leads to the
observational richness of QFT. Once we leave this setting by going to curved
spacetime or to QFT in KMS thermal representations, or if we restrict a
Minkowski spacetime theory to a Rindler wedge with the Hamiltonian being now
the boost operator with its two-sided spectrum, in all these cases we are
loosing not only the setting of scattering theory but also the very notion of
particles as elementary systems with respect to the Poincar\'{e} group. With
it also most of the observational wealth related to scattering theory is lost.
Any deviation from Poincar\'{e} covariance also endangers the existence of a
vacuum. The restriction to the Rindler world preserves the Fock space particle
structure of the free field Minkowski QFT, but it looses its intrinsic
physical significance with respect to the Rindler situation\footnote{There is
of course the mathematical possibility of choosing a groundstate
representation for a Rindler world instead of restricting the Minkowski vacuum
and to have a finite number of "quanta" (excitations). But there is no reason
for believing that these objects fall into the range of validity of the
Haag-Ruelle scattering theory which is the hallmark of particle physics as we
know it.}. Since the Minkowski vacuum restricted to the Rindler world is now a
thermal KMS state, there is no particle scattering theory in the "boost time"
in such a thermal situation. The remaining observable phenomena are
Hawking-like \cite{Haw} radiation densities and their fluctuations i.e.
observables such as they are presently studied in the cosmic background
radiation. Some of the conceptual problems related to the Unruh effect
\cite{Un} have been addressed in the philosophically oriented literature
\cite{Cl-Hal}\cite{Arag}. Quantum fields are not directly accessible to
measurements\footnote{An opposing opinion to this "interpolating field point
of view" can be found in \cite{Wald2}.} and therefore the problem what happens
to the wealth of particle physics in such QFT requires more research.

Formally the local covariance principle forces the construction of a QFT on
all causally complete manifolds and their submanifolds at once. So the QFT in
Minkowski spacetime with its particle interpretation is always part of the
solution. What one would like to have is a more direct physical connection
e.g. a particle concept in the tangent space or something in this direction.

The conceptual differences between a DPI relativistic QM and QFT are enormous,
but in order to perceive this, one has to get away from the shared properties
of the quantization formalism, a step which with the exception of \cite{Haag}
is usually not undertaken in textbooks whose prime objective is to get to
calculational recipes with the least conceptual investment. It is the main
purpose of the following sections to highlight these contrasts by going more
deeply into QFT.

There are certain folkloric statements about the relation QM--QFT whose
refutation does not require much conceptual sophistication. For example in
trying to make QFT more susceptive to newcomers it is sometimes said that a
free field is nothing more than a collection of infinitely many coupled
oscillators. Although not outright wrong, this characterization misses the
most important property of how spacetime enters as an ordering principle into
QFT. It would not help any newcomer who knows the quantum oscillator, but has
not met a free field before, to construct a free field from such a verbal
description. Even if he manages to write down the formula of the free field he
would still have to appreciate that the most important aspect is the causal
localization and not that what oscillates. This is somewhat reminiscent of the
alleged virtue from equating QM via Schr\"{o}dinger's formulation with
classical wave theory. What may be gained for a newcomer by appealing to his
computational abilities acquired in classical electrodynamics, is more than
lost in the conceptual problems which he confronts later when facing the
subtleties of entanglement in quantum physics.

\section{More on Born versus covariant localization}

In this section it will be shown that the difference between QM and LQP in
terms of their localization result in a surprising distinction in their notion
of entanglement. We will continue to use the word \textit{Born localization}
for the probability density of the x-space Schroedinger wave function
$p(x)=\left\vert \psi(x)\right\vert ^{2};$ whereas its adaptation to the
invariant inner product of relativistic wave functions which was done by
Newton and Wigner \cite{N-W} and will referred to as BNW localization. Being a
bona fide probability density, one may characterize the BNW localization in a
spatial region $R\in\mathbb{R}^{3}$ at a given time in terms of a
\ localization projector $P(R)$ which appears in the spectral decomposition of
the selfadjoint position operator. The standard version of QM and the various
settings of measurement theory rely heavily on these projectors; without BNW
localization and the ensuing projectors it would be impossible to formulate
the conceptual basis for the time-dependent scattering theory of QM and QFT.

The BNW position operator and its family of spatial region-dependent
projectors $P(R)$ is not covariant under Lorentz boosts. For Wigner, to whom
modular localization was not available, this frame dependence raised doubts
about the conceptual soundness of QFT. Apparently the existence of completely
covariant correlation functions in renormalized perturbation theory did not
satisfy him, he wanted an understanding from first principles and not as an
outgrowth of some formalism.

The lack of covariance of BNW localization in finite time propagation leads to
frame-dependence and superluminal effects, which is why the terminology
"relativistic QM" has to be taken with a grain of salt. However, as already
emphasized, in the asymptotic limit of large timelike separation as required
in scattering theory, the covariance, frame-independence and causal relations
are recovered. As shown in section 3 one obtains a Poincar\'{e}-invariant
unitary M\o ller operator and S-matrix whose DPI construction within an
interacting n-particle Wigner representation of the Poincar\'{e} group which
also guaranties the validity of all the macro-causality requirements
(spacelike clustering, absence of timelike precursors, causal rescattering)
which can be formulated in a particle setting i.e. without taking recourse to
interpolating local fields. Even though the localizations of the individual
particles are frame-dependent, the asymptotic relation between BNW-localized
events is given in terms of the geometrically associated \textit{covariant
on-shell momenta} or 4-velocities which describe the asymptotic movement of
the c.m. of wave packets. In fact \textit{all observations on particles always
involve BNW localization} measurements.

The situation of propagation of DPI is similar to that of propagation of
acoustic waves in an elastic medium; although in neither case there is a
limiting velocity, there exists a maximal "effective" velocity, for DPI this
is c and in the acoustic case this is the velocity of sound in the particular medium.

In comparing QM with QFT it is often convenient in discussions about
conceptual issues to rephrase the content of (nonrelativistic) QM in terms of
operator algebras and states (in the sense of positive expectation functional
on operator algebras); in this way one also achieves more similarity with the
formalism of QFT and develops a greater awareness for genuine conceptual
antinomies. In this Fock space setting the basic quantum mechanical operators
are the creation/annihilation operators $a^{\#}(\mathbf{x})$ with%
\begin{equation}
\left[  a(\mathbf{x}),a^{\ast}(\mathbf{y})\right]  _{grad}=\delta
(\mathbf{x}-\mathbf{y}) \label{qu}%
\end{equation}
where for Fermions the graded commutator stands for the anticommutator. In the
QFT setting it is not forbidden to work with such operators (the Fourier
transforms of the Wigner creation/annihilation operators), except that it
becomes nearly impossible to keep track of covariance and express local
observables in terms of them\footnote{In fact local observables would appear
nonlocal. The incorrect use of these operators led Irving Segal to the
conclusion that local observable subalgebras in QFT are quantum mechanical
type I factors a claim which he withdrew after becoming aware of the results
by Araki \cite{Ara} who showed that they are of type III (later refined to the
unique "hyperfinite type III$_{1}"$).}.

The ground state for T=0 zero matter density states is annihilated by $a(x),$
whereas for finite density one encounters a state in which the levels are
occupied up to the Fermi surface in case of Fermions, and contains a
Bose-Einstein condensate groundstate in case of Bosons.

In QFT the identification of pure states with state-vectors of a Hilbert space
has no intrinsic meaning and often cannot be maintained in concrete
situations. For the reason of facilitating the comparison with QM we use the
unified Fock space setting instead of the Schroedinger formulation. Although
DPI is formulated in Fock space, there is no useful second quantized formalism
(\ref{qu}).

The global algebra which contains all observables independent of their
localization is the algebra $B(H)$ of all bounded operators in Hilbert space.
Physically important unbounded operators are not members but rather have the
mathematical status of being affiliated with $B(H)$ and its subalgebras; this
bookkeeping makes it possible to apply powerful theorems from the theory of
operator algebras (whereas unbounded operators are treated on a case to case
basis). $B(H)$ is the correct global description whenever the physical system
under discussion arises as the weak closure of a ground state representation
of an irreducible system of operators\footnote{The closure in a thermal
equilibrium state associated with a continuous spectrum Hamiltonian leads to a
unitarily inequivalent (type III) operator algebra without minimal
projectors.} be it QM or LQP$.$ According to the classification of operator
algebras, $B(H)$ and all its multiples are of Murray von Neumann type
$I_{\infty}$ whose characteristic property is the existence of minimal
projectors; in the irreducible case these are the one-dimensional projectors
belonging to measurements which cannot be refined$.$ There are prominent
physical states which lead to different global situations as e.g. thermal KMS
states, but for the time being our interest is in ground states.

The structural differences between QM and LQP emerge as soon as one defines a
physical substructure on the basis of localization. It is well known that a
dissection of space into nonoverlapping spatial regions i.e. $\mathbb{R}%
^{3}=\cup_{i}R_{i}$ implies via Born localization a tensor factorization of
$B(H)$ and $H$%
\begin{align}
&  B(H)=%
{\displaystyle\bigotimes\limits_{i}}
B(H(R_{i}))\\
&  H=%
{\displaystyle\bigotimes_{i}}
H(R_{i}),~P(R_{i})H=H(R_{i})\nonumber\\
&  \mathbf{\vec{X}}_{op}=\int a^{\ast}(\vec{x})\vec{x}a(\vec{x})d^{3}%
x=\int\vec{x}dP(\vec{x})
\end{align}
where the third line contains the definition of the position operator and its
spectral decomposition in the bosonic Fock space. Hence there is orthogonality
between subspaces belonging to localizations in nonoverlapping regions
(orthogonal Born projectors) and one may talk about states which are pure in
$H(R_{i}).$ As well known from the discussion of entanglement, a pure state in
the global algebra $B(H)$ may not be of the special tensor product form but
rather be a superposition of factorizing states; the Schmidt decomposition is
a method to achieve this with an intrinsically determined basis in the case of
a bipartite tensor factorization.

States which are not tensor products, but rather superpositions of such, are
called entangled; their reduced density matrix obtained by averaging over the
environment of $R_{i}$ describes a mixed state on $B(H(R_{i}))$. This is the
standard formulation of QM in which pure states are vectors and mixed states
are density matrices.

Although this quantum mechanical entanglement can be related to the notion of
entropy, it is an entropy in the sense of \textit{information theory} and not
in the \textit{thermal sense}. One cannot create a physical temperature as a
quantitative measure of the degree of quantum mechanical entanglement in this
way. which results from BNW-restricting pure global states to a finite region
and its outside environment. In particular the ground state always factorizes,
a spatial tensor factorization never causes vacuum polarization and
entanglement in QM setting. The net structure of $B(H)$ in terms the
subalgebras $B(H(R_{i})$) is of a kinematical kind; although the reduced state
may be impure, there is no $B(H(R)$ reduced Hamiltonian relative to which an
impure state in QM becomes a KMS state. Here QM stands for any QT without a
maximal propagation speed i.e. one which lacks causal propagation and vacuum polarization.

The LQP counterpart of the Born-localized subalgebras at a fixed time are the
observable algebras $\mathcal{A(O)}$ for spacetime double cone regions
$\mathcal{O}$ obtained from spatial regions $R$ by causal completion
$\mathcal{O=}R^{\prime\prime}$ (causal complement taken twice); they form what
is called in the terminology of LQP a \textit{local net} \{$\mathcal{A(O)}%
$\}$_{\mathcal{O}\subset M}$ of operator algebras indexed by regions in
Minkowski spacetime $\cup\mathcal{O=M}$ which is subject to the natural and
obvious requirements of isotony ($\mathcal{A(O}_{1}\mathcal{)}\subset
\mathcal{A(O}_{2}\mathcal{)}$ if $O_{1}\subset O_{2}$) and causal locality,
i.e. the algebras commute for spacelike separated regions.

The connection with the standard formulation of QFT in terms of pointlike
fields is that smeared fields $\Phi(f)=\int\Phi(x)f(x)d^{4}x$ with
$suppf\subset\mathcal{O}~$under reasonable general conditions generate local
algebras. Pointlike fields, which by themselves are too singular to be
operators (even if admitting unboundedness), have a well-defined mathematical
meaning as operator-valued distributions briefly referred to as generators of
algebras. The singular nature of generating fields is therefore not a
pathological aspect leading to inescapable ultraviolet catastrophes, but
rather a natural attribute of passing from classical to quantum fields.

The real cumbersome aspect is not their singular behavior but their multitude;
there are myriads of fields which generate the same net of local operator
algebras and interpolate the same particles whereas in classical field theory
they could be distinguished by classical field measurements.

In this sense generating fields play a similar role in LQP as coordinates in
modern differential geometry i.e. they coordinatize the net of spacetime
indexed operator algebras and only the latter has an intrinsic meaning; in
particular the particles and their collision theory can be obtained from the
local net without being forced to distinguish individual operators within a
local algebra. But as the use of particular coordinates often facilitates
geometrical calculations, the\textbf{ }use of particular fields, with e.g. the
one with the lowest short-distance dimension within the infinite charge
equivalence class of fields, can greatly simplify\footnote{The field which is
"basic" in the sense of a Lagrangian field in a Lagrangian approach is
generally simpler to deal with than composites of that fields (the Massive
Thirring field is simpler than the Sine-Gordon field which maybe derived from
it).} calculations in QFT. Therefore it is a problem of practical importance
to construct a covariant basis of locally covariant pointlike fields of an
equivalence class.

For massive free fields and for massless free fields of finite helicity such a
basis is especially simple; the "Wick-basis" of composite fields still follows
in part the logic of classical composites (apart from the definition of the
double dot : :). This remains so even in the presence of interactions in which
case the Wick-ordering gets replaced by the technically more demanding "normal
ordering" \cite{Zi}. For free fields in curved spacetime (CST) and the
definition of their composites it is important to require the \textit{local
covariant transformation behavior} under local isometries \cite{Ho-Wa2}. The
conceptual framework of QFT in CST in the presence of \textit{interactions}
has also been largely understood \cite{BFV}.

We now return to the main question namely: what changes if we pass from the
BNW localization of QM/DPI to the causal localization of LQP? The crucial
property is that a localized algebra $\mathcal{A(O)}\subset B(H)$ together
with its commutant $\mathcal{A(O)}^{\prime}$ (which under very general
conditions\footnote{In fact this duality relation can always be achieved by a
process of maximalization (Haag dualization) which increases the degrees of
freedom inside $\mathcal{O}.$ A pedagogical illustration based on the
"generalized free field" can be found in \cite{Du-Re}.} is equal to the
algebra of the causal disjoint of $\mathcal{O}$ i.e. $\mathcal{A(O)}^{\prime
}=\mathcal{A(O}^{\prime}\mathcal{))}$ are two von Neumann factor algebras i.e.%
\begin{equation}
B(H)=\mathcal{A(O)\vee A(O)}^{\prime},\text{ }\mathcal{A(O)\cap A(O)}^{\prime
}=\mathbb{C}\mathbf{1} \label{loc}%
\end{equation}
In contrast to the QM algebras the local factor algebras are not of type I and
$B(H)$ does \textit{not tensor-factorize} in terms of them, in fact they
cannot even be embedded into a $B(H_{1})\otimes B(H_{2})$ tensor product. The
prize to pay for ignoring this important fact and imposing wrong structures is
the appearance of spurious \textit{ultraviolet divergences,} the typical way
of a QFT model\ to resist enforcing an incompatible structure on it.

On the positive side, as will be explained in the second part of this essay,
without this significant change in the nature of algebras there would be no
holography onto causal horizons and the resulting huge symmetry enhancement to
infinite-dimensional (BMS) groups, and of course there would be no thermal
behavior caused by localization and a fortiori no area-proportional
localization entropy.

The situation in LQP is radically different from that of entanglement and pure
versus mixed states in QM since local algebras as $\mathcal{A(O})$ have
\textit{no pure states at all}; so the dichotomy between pure and mixed states
breaks down and the kind of entanglement caused by field theoretic
localization is much more violent then that coming from
BNW-localization\footnote{By introducing in addtion to free fields $A(x)$
which are covariant Fourier transforms also noncovariant Fourier transforms
$a(\vec{x},t),a^{\ast}(\vec{x},t)$ one can explicitly that the latter are
relatively nonlocal.}, in the terminology of Ruetsche \cite{Rue} these states
are \textit{instrinsically mixed.} This implies that the standard pure-mixed
dichotomy does not extend beyond QM i.e. such intrinsically mixed states do
not exist in any natural way on $B(H).$ At the moment in which they come into
being as e.g. thermodynamic limit states in the infinite volume limit, the
algebra has ceased to be of the quantum mechanical B(H) type and become a type
III operator algebra \cite{Robin}. The thermodynamic limit construction at
finite temperature gives also the correct hint to the nature of intrinsically
mixed states; they are typically "singular" KMS states i.e. KMS states which
although being the thermodynamic limits of Gibbs state cannot themselves be
represented in the Gibbs form because the KMS Hamiltonian has continuous spectrum.

Unlike Born localization, causal localization is not related to position
operators and projectors $P(R);$ rather the operator algebras $\mathcal{A(O})$
are of an entirely different kind than those met in ground state (zero
temperature) QM ; they are all isomorphic to one abstract object, the
hyperfinite type III$_{1}$ von Neumann factor also referred to as \textit{the
monad} the unique factor behind Araki's 1963 discovery \cite{Ara}\textit{.} As
will be seen later LQP creates its wealthy mansion from just this one kind of
brick; all its structural richness comes from positioning the bricks, there is
nothing hidden in the structure of one bricks. In a later section it will be
explained how this emerges from modular localization and a related operator formalism.

The situation does not change if one takes for $\mathcal{O}$ a region $R$ at a
fixed time; as stated before, in a theory with finite propagation speed one
has$\mathcal{A}(R)=\mathcal{A}(D(R)),$ where $D(R)$ is the diamond shaped
double cone subtended by $R$ (the causal shadow of $R$)$.$ Even if there are
no pointlike generators and if the theory (as the result of the existence of
an elementary length) only admits a macroscopically localized net of algebras
(e.g. a net of non-trivial wedge-localized factor algebras $\mathcal{A}(W)$
with trivial double cone intersection algebras $\mathcal{A(O})=\{c\mathbf{1}%
\}),$ the algebras would still not tensor factorize $B(H)\neq\mathcal{A}%
(W)\otimes\mathcal{A}(W^{\prime}).$ Hence the properties under discussion are
not directly related to the presence of singular generating
pointlike/stringlike fields but are connected to the existence of well-defined
(sharp) causal shadows. There is a hidden singular aspect in the sharpness of
the $\mathcal{O}$-localization which generates infinitely large vacuum
polarization clouds on the causal horizon of the localization. In the last
section a method (splitting) will be presented which permits to define a
split-distance dependent, but otherwise intrinsically defined finite thermal entropy.

Most divergencies (but not all, since the divergence of localization entropy
for vanishing splitting distance is an unavoidable consequence of the
principles) in QFT are the result of conceptual errors in the formulation
resulting from tacitly identifying QFT with some sort of relativistic
QM\footnote{The correct treatment of perturbation theory which takes into
account the singular nature of pointlike quantum fields may yield more free
parameters than in the classical setting, but one is never required to
confront infinities or cut-offs.} and in this way ignoring the intrinsically
singular nature of pointlike localized fields.

Often it is thought that the avoidance of locality in favor of
\textit{nonlocal} covariant operators eliminates the singular short distance
behavior. But this is not quite true as evidenced by the Kallen-Lehmann
representation of a covariant scalar object%
\begin{equation}
\left\langle A(x)A(y)\right\rangle =\int\Delta_{+}(x-y,\kappa^{2})\rho
(\kappa^{2})d\kappa^{2}%
\end{equation}
which was proposed precisely to show that even without demanding locality, but
retaining only covariance and the Hilbert space structure (positivity), a
certain singular behavior of covariant objects is unavoidable. In the DPI
scheme this was avoided, because even though there are particles at all times,
there are no covariant (tensors, spinors) objects at finite times, the only
covariant quantity arises in the form of the invariant S-matrix in the
$t\rightarrow\infty$ $\ $limit. The next section shows that a separation
between covariance and localization in the pursuit of a less singular more
nonlocal theory is a futile endeavour, at least as long as one does not
subject spacetime itself to a radical revision.

In the algebraic formulation the covariance requirement refers to the geometry
of the localization region $\mathcal{A}(\mathcal{O})$ i.e.%
\begin{equation}
U(a,\Lambda)\mathcal{A(O})U(a,\Lambda)^{\ast}=\mathcal{A}(\mathcal{O}%
_{a,\Lambda})
\end{equation}
whereas no additional requirement about the transformation behavior under
finite dimensional (tensor, spinor) Lorentz representations (which would bring
back the unboundedness and thus prevent the use of powerful theorems in
operator algebras) is imposed for the individual operators. The singular
nature of pointlike generators (if they exist) is then a purely mathematical
consequence. Using such singular objects in pointlike interactions in the same
way as one uses operators in QM leads to self-inflicted divergence problems.

We have seen that although QM and QFT can be described under a common
mathematical roof of $C^{\ast}$-algebras with a state functional, as soon as
one introduces the physically important localization structure, significant
conceptual differences appear. These differences show up in the presence of
vacuum polarization in QFT as a result of causal localization and they tend to
have dramatic consequences; the most prominent ones will be presented in this
and the subsequent sections, more will be contained in the second essay.

The net structure of the observables allows a \textit{local comparison of
states}: two states are locally equal in a region $\mathcal{O}$ if and only if
the expectation values of all operators in are the same in both states. Local
deviations from any state, in particular from the vacuum state, can be
measured in this manner; states which are equal on the causal complement
$\mathcal{A(O}^{\prime}\mathcal{)}$ that are indistinguishable from the vacuum
are called localizable in $\mathcal{A(O)}$ ("strictly localized states" in the
sense of Licht \cite{Licht}) can be defined. Due to the unavoidable
correlations in the vacuum state in relativistic quantum theory (the
Reeh-Schlieder property \cite{Haag}), the space $H(\mathcal{O})$ obtained by
applying the operators in $\mathcal{A(O)}$ to the vacuum is, for any open
region $\mathcal{O}$, dense in the Hilbert space and thus far from being
orthogonal to $H(\mathcal{O}^{\prime})$. This somewhat counter-intuitive fact
is inseparably linked with a structural difference between the local algebras
and the algebras encountered in non-relativistic quantum mechanics (or the
global algebra of a quantum field associated with the entire Minkowski
space-time) as mentioned in connection with the breakdown of
tensor-factorization (\ref{loc}).

The result is a particular benevolent form of "Murphy's law" for interacting
QFT: \textit{everything which is not forbidden (by superselection rules) to
couple, really does couple}. On the level of interacting particles this has
been termed \textit{nuclear democracy}: any particle whose superselected
charge is contained in the spectrum which results from fusing the charges in a
cluster of particles can be viewed as a bound state of that cluster of
particles. Nuclear democracy even strips a particle with a fundamental charge
of its individuality since such an object can be considered as bound of itself
+ an arbitrary number of particles with non-fundamental charges. This renders
interacting QFT conceptually much more attractive and fundamental than QM, but
it also contributes to its computational complexity i.e. the benevolent
character of Murphy's LQP law unfortunately does not necessarily extend to the
computational side, at least if one limits oneself to the standard tools of QT.

The Reeh-Schlieder property \cite{Haag} (in more popular but less precise
terminology: the "state-field relation") is perhaps the strongest realization
of Murphy's law since it secures the existence of a localization region
dependent dense subspace $H(\mathcal{O})=\mathcal{A(O})\Omega\subset H$ which
cannot be associated with a nontrivial projector. It also implies that the
expectation value of a projection operator localized in a bounded region
\textit{cannot} be interpreted as the probability of detecting a particle-like
object in that region, since it is necessarily nonzero if acting on the vacuum
state. The $\mathcal{A}(\mathcal{O})$-reduced ground state is a KMS thermal
state at a appropriately normalized (Hawking) temperature (more in part II).
The intrinsically defined \textit{modular "Hamiltonian" } associated via
modular operator theory\footnote{The modular Hamitonian is the infinitesimal
generator $K_{\operatorname{mod}}$ of the modular group $\Delta^{it}\equiv
e^{-itK_{\operatorname{mod}}}.$(see next two sections).} to a "standard pair"
($\mathcal{A(O)},\Omega_{vac}$) is always available in the mathematical sense
but allows a physical interpretation only in those rare cases when there
exists an invariance group of $\mathcal{O}$ which is a subgroup of the
spacetime group leaving $\Omega_{vac}$ invariant. Well known cases are the
Lorentz boost for the wedge region in Minkowski spacetime (the Unruh effect)
and the generator of a double-cone preserving conformal transformation in a
conformal theory and certain Killing symmetries in black hole physics. Its
general purpose is to give an intrinsic description of the $\mathcal{A}%
(\mathcal{O}$)- reduced vacuum state in terms of an KMS state of an Hamiltoian
"movement" where we used brackets in order to highlight the fact that this is
generally not a geometric movement but only an algebraic automorphism of
$\mathcal{A}(\mathcal{O}$) (and simultaneously of $\mathcal{A}(\mathcal{O}%
^{\prime}$)) which respects the geometric boundaries (the causal horizon) of
$\mathcal{O}$ \footnote{In fact it induces a geometric movement \textit{on}
the horizon.}. It is never the Hamiltonian associated with a globally inertial
reference frame as in case of heat bath thermal systems.

There exists in fact a whole family \textit{of modular Hamiltonians} since the
operators in $\mathcal{A(O)}$ naturally fulfill the KMS condition for any
standard pair ($\mathcal{A(\check{O})},\Omega_{vac}$) for $\mathcal{\check
{O}\supset O}$: i.e. the different modular Hamiltonians and the KMS states
change with the causally closed world $\mathcal{\check{O}}$ of the observer.
The surprising aspect is that the causal localization structure of one QFT
leads to an infinite supply of different Hamiltonians without any change of
interactions. The change of the modular Hamiltonian $K_{\mathcal{O}}$ via a
change of the localization region will lead to a new Hamiltonian whose
automorphic movement maintains the new region but leaves (after some "modular
time") the old region i.e. this is not a family of Hamiltonians on a quantum
mechanical algebra. Of particular interest is the restriction of a modular
automorphism to the horizon of a causally closed region $Hor(\mathcal{O});$
there are good indications that this defines a diffeomorphism which belongs to
the infinite dimensional Bondi-Metzner-Sachs subgroup of a gigantic symmetry
group of holographic projection onto horizons (see part II on holography).

The situation just described is one of extreme "virtuality", i.e. there is
generally not even the possibility to view it in terms of an
Gedankenexperiment of a non-inertial (accelerated) $\mathcal{O}$-confined
observer for whom the modular movement is an $\mathcal{O}$ - preserving
diffeomorphism; such pure algebraic movements without individual orbits are
often called "fuzzy". Whenever the modular movement passes to a diffeomorphism
one can at least envisage a Gedankenexperiment which keeps the observer on an
$\mathcal{O}$-preserving track by appropriate accelerations. The only
geometric case in Minkowski spacetime is the situation proposed first by Unruh
\cite{Un}, when $\mathcal{O}$ is a wedge i.e. a region $W$ which is bounded by
two intersecting lightfronts which only share the 2-dim. edge of their
intersection.. Conformal theories for which the observables live in the
Dirac-Weyl compactification $\widetilde{M}$ of the Minkowski spacetime lead to
modular diffeomorphisms even for compact double cones $\mathcal{D}%
$\footnote{The region obtained by intersecting a forward lightcone with
arbitrary apex \ with an backward lightcone;}.

The most interesting and prominent case comes about when spacetime curvature
is creating a black hole. In case there are time-like Killing orbits and an
extension of the spacetime such that the black hole horizon is a event horizon
in the sense of dividing the extended manifold into a causally inside/outside
with separate Killing movements, one is in the classical Hawking-like
situation. What one in additions needs for the quantum setting is the
existence of a quantum state which is invariant under the Killing group action.

In the case of the Schwarzschild black hole all these requirements are
fulfilled, the extension is the Schwarzschild-Kruskal extension and the
invariant state is the Hartle-Hawking state $\Omega_{H-H}$. \ In this case
($\mathcal{A(O}_{S-K}),\Omega_{H-H}$) is a standard pair and the modular
movement is the Killing orbit which respects the black hole event horizon.
Whereas the causal horizons in the previous Minkowski spacetime examples was
an extremely "fleeting" object, a black hole event horizon has an intrinsic
metric-imprinted position. Besides their astrophysical interests, black holes
are therefore of considerable philosophical interest. The only future
development which could still enforce a significant modification of the
present concepts is the still unknown \textit{quantum gravity} (more remarks
on QG in part II).

For computations of thermal properties, including thermal entropy, it does not
matter whether the horizon is a "fleeting" observer-dependent causal
localization\footnote{The localization entropy which depends on the "split"
size (see below) is however an important property of the model, even if it not
directy experimentally accessible.} horizon or a fixed curvature generated
black hole \textit{event horizon}; only its direct observable significance
depends on the black hole event horizon. This leads to a picture about the
LQP-QG (quantum gravity) interface which is somewhat different from that in
most of the literature; we will return to these issues in connection with the
presentation of the split property in part II of the essay.

Causality in relativistic quantum field theory is mathematically expressed
through local commutativity, i.e., mutual commutativity of the algebras
$\mathcal{A(O)}$ and $\mathcal{A(O}^{\prime}$). There is an intimate
connection of this property with the possibility of preparing states that
exhibit no mutual correlations for a given pair of causally disjoint regions.
In fact, in a recent paper Buchholz and Summers \cite{Bu-Su} show that local
commutativity is a necessary condition for the existence of such uncorrelated states.

Conversely, in combination with some further properties related to degrees of
freedom densities (split property \cite{Do-Lo}, existence of scaling limits
\cite{Bu-Ve}), local commutativity leads to a very satisfactory picture of
\textit{statistical independence} and \textit{local preparabilty of states} in
relativistic quantum field theory. We refer to \cite{Sum}\cite{Wer} for
thorough discussions of these matters and \cite{Yngva}\cite{MSY} for a brief
review of some physical consequences. The last two papers also explain how the
above mentioned concepts avoids spurious problems rooted in assumptions that
are in conflict with basic principles of relativistic quantum physics. In
particular it can be shown how an alleged difficulty \cite{Heger}\cite{Bu-Yn}
with Fermi's famous Gedankenexperiment \cite{Fermi}, which Fermi proposed in
order to show that the velocity of light is also the limiting propagation
velocity in quantum electrodynamics, can be resolved by taking \cite{Yngva}
into account the progress on the conceptual issues of causal localization and
the gain in mathematical rigor since the times of Fermi.

After having discussed some significant conceptual differences between QM and
LQP, one naturally asks for an argument why and in which way QM appears as a
nonrelativistic limit of LQP. The standard kinematical reasoning of the
textbooks is acceptable for fermionic/bosonic systems in the sense of "FAPP",
but has not much strength on the conceptual level. To see its weakness,
imagine for a moment that we would live in a 3-dim. world of anyons (abelian
plektons, where plektons are Wigner particles with braid group statistics).
Such relativistic objects are by their very statistics so tightly interwoven
that there simply are no compactly localized free fields which only create a
localized anyon without a vacuum polarization cloud admixture. In such a world
no nonrelativistic limit which maintains the spin-statistic connection could
lead to QM, the limiting theory would rather \textit{remain a nonrelativistic
QFT}. In order to avoid misunderstandings, its is not claimed here that the
issue of nonrelativistic limits of any interacting relativistic QFT is
mathematically understood\footnote{The arguments about the nonrelativistic
limit og QFT have remained metaphoric; however the existence of exactly solved
interacting 2-dim. QFTs raises now hopes that age old problem will be better
understood.}, rather the statement is that plektonic (braid-group) commutation
relations, relativistic or nonrelativistic, interacting or not, are
incompatible with the structure of (Schroedinger) QM. In 4-dimensional
spacetime there is no such obstacle against QM, simply because it is not the
Fermi/Bose statistics which causes vacuum polarization; to formulate it more
provocatively: there would be no Schroedinger QM without the existence of free
relativistic fermions/bosons.

\section{Modular localization}

Previously it was mentioned on several occasions that the localization
underlying QFT can be freed from the contingencies of field coordinatizations.
This is achieved by a physically as well mathematically impressive, but for
historic and sociological reasons little known theory. Its name "modular
theory" is of mathematical origin and refers to a vast generalization of the
(uni)modularity encountered in the relation between left/right Haar measure in
group representation theory. In the middle 60s the mathematician Tomita
presented a significant generalization of this theory to operator algebras and
in the subsequent years this theory received essential improvements from
Takesaki and later from Connes.

At the same time Haag, Hugenholtz and Winnink published their work on
statistical mechanics of open systems \cite{Haag}. When the physicists and
mathematicians met at a conference in Baton Rouge in 1966, there was surprise
about the similarity of concepts, followed by deep appreciation about the
perfection with which these independent developments supported each other
\cite{Borch}. Physicists not only adapted mathematical terminology, but
mathematicians also took some of their terminology from physicists as e.g.
\textit{KMS states} which refer to Kubo, Martin and Schwinger who introduced
an analytic property of Gibbs states merely as a computational tool (in order
to avoid computing traces), Haag, Hugenholtz and Winnink realized that this
property (which they termed the KMS property) is the only aspect which
survives in the thermodynamic limit when the trace formulas loses its meaning
and must be replaced by the analytic KMS boundary condition.

This turned out to be the right concept for formulating and solving problems
directly in the setting of open systems. In the present work the terminology
is mainly used for thermal states of open systems which are not Gibbs states.
They are typical for LQP, for example every multiparticle state $\Omega
_{particle}$ of finite energy, including the vacuum, (i.e. every physical
particle state) upon restriction to a local algebra $\mathcal{A(O})$ becomes a
KMS state with respect to a "modular Hamiltonian" which is canonically
determined by ($\mathcal{A(O}),\Omega_{particle}$).

Connes, in his path-breaking work on the classification of von Neumann factors
\cite{Connes}, made full use of this hybrid math.-phys. terminology which
developed after Baton Rouge. Nowadays one can meet mathematicians who use the
KMS property but do not know that this was a mere computational tool by 3
physicists (Kubo. Martin and Schwinger) to avoid calculating traces and that
the conceptual aspect was only realized later by Haag Hugenholtz and Winnink
who gave it its final name. One can hardly think of any other confluence of
mathematical and physical ideas on such a profound and at the same time equal
and natural level as in modular theory; even the Hilbert space formalism of QM
already existed for many years before quantum theorists became aware of its use.

About 10 years after Baton Rouge, Bisognano and Wichmann \cite{Bi-Wi}
discovered that a vacuum state restricted to a wedge-localized operator
algebra $\mathcal{A(}W)$ in QFT defines a modular setting in which the
restricted vacuum becomes a thermal KMS state with respect to the
wedge-affiliated L-boost "Hamiltonian". This step marks the beginning of a
very natural yet unexpected relation between thermal and geometric properties,
one which is totally characteristic for QFT i.e. which is not shared by
classical theory nor by QM. Thermal aspects of black holes were however
discovered independent of this work, and the first physicist who saw the
connection with modular theory was Geoffrey Sewell \cite{Sew}.

The theory becomes more accessible for physicists if one introduces it first
in its more limited spatial- instead of its full algebraic- context. Since as
a foundational structure of LQP it merits more attention than it hitherto
received from the particle physics community, some of its methods and
achievements will be presented in the sequel.

It has been realized by Brunetti, Guido and Longo\footnote{In a more limited
context and with less mathematical rigor this was independently proposed in
\cite{Sch}.} \cite{BGL} that there exists a natural localization structure on
the Wigner representation space for any positive energy representation of the
proper Poincar\'{e} group. The starting point is an irreducible representation
$U_{1}$of the Poincar\'{e}%
\'{}%
group on a Hilbert space $H_{1}$ that after "second quantization" becomes the
single-particle subspace of the Hilbert space (Wigner-Fock-space) $H_{WF}$ of
the quantum fields act\footnote{The construction works for arbitrary positive
energy representations, not only irreducible ones.}. In the bosonic case the
construction then proceeds according to the following steps \cite{BGL}%
\cite{Fa-Sc}\cite{MSY}.

One first fixes a reference wedge region, e.g. $W_{0}=\{x\in\mathbb{R}%
^{d},x^{d-1}>\left\vert x^{0}\right\vert \}$ and considers the one-parametric
L-boost group (the hyperbolic rotation by $\chi$ in the $x^{d-1}-x^{0}$ plane)
which leaves $W_{0}$ invariant; one also needs the reflection $j_{W_{0}}$
across the edge of the wedge (i.e. along the coordinates $x^{d-1}-x^{0}$). The
$\mathfrak{j}_{W_{0}}$ extended Wigner representation is then used to define
two commuting wedge-affiliated operators%
\begin{equation}
\mathfrak{\delta}_{W_{0}}^{it}=\mathfrak{u}(0,\Lambda_{W_{0}}(\chi=-2\pi
t)),~\mathfrak{j}_{W_{0}}=\mathfrak{u}(0,j_{W_{0}})
\end{equation}
where attention should be paid to the fact that in a positive energy
representation any operator which inverts time is necessarily
antilinear\footnote{The wedge reflection $\mathfrak{j}_{W_{0}}$ differs from
the TCP operator only by a $\pi$-rotation around the W$_{0}$ axis.}. A unitary
one- parametric strongly continuous subgroup as $\delta_{W_{0}}^{it}$ can be
written in terms of a selfadjoint generator $K$ as $\delta_{W_{0}}%
^{it}=e^{-itK_{W_{0}}}$ and therefore permits an "analytic continuation" in
$t$ to an unbounded densely defined positive operators $\delta_{W_{0}}^{s}$.
With the help of this operator one defines the unbounded antilinear operator
which has the same dense domain as its "radial" part%
\begin{equation}
\mathfrak{s}_{W_{0}}=\mathfrak{j}_{W_{0}}\mathfrak{\delta}_{W_{0}}^{\frac
{1}{2}},\text{ }\mathfrak{j\delta}^{\frac{1}{2}}\mathfrak{j}\mathfrak{=\delta
}^{-\frac{1}{2}} \label{c.r.}%
\end{equation}

Whereas the unitary operator $\delta_{W_{0}}^{it}$ commutes with the
reflection, the antiunitarity of the reflection changes the sign in the
analytic continuation which leads the commutation relation between $\delta$
and $\mathfrak{j}$ in (\ref{c.r.}). This causes the involutivity of the
s-operator on its domain, as well as the identity of its range with its
domain
\begin{align*}
\mathfrak{s}_{W_{0}}^{2}  &  \subset\mathbf{1}\\
dom~\mathfrak{s}  &  =ran~\mathfrak{s}%
\end{align*}
Such operators which \textit{are unbounded and yet involutive} on their domain
are very unusual; according to my best knowledge they only appear in modular
theory and it is precisely these unusual properties which are capable to
encode geometric localization properties into domain properties of abstract
quantum operators, a fantastic achievement completely unknown in QM. The more
general algebraic context in which Tomita discovered modular theory will be
mentioned later.

The idempotency means that the s-operator has $\pm1$ eigenspaces; since it is
antilinear, the +space multiplied with $i$ changes the sign and becomes the -
space; hence it suffices to introduce a notation for just one eigenspace%
\begin{align}
\mathfrak{K}(W_{0})  &  =\{domain~of~\Delta_{W_{0}}^{\frac{1}{2}%
},~\mathfrak{s}_{W_{0}}\psi=\psi\}\\
\mathfrak{j}_{W_{0}}\mathfrak{K}(W_{0})  &  =\mathfrak{K}(W_{0}^{\prime
})=\mathfrak{K}(W_{0})^{\prime},\text{ }duality\nonumber\\
\overline{\mathfrak{K}(W_{0})+i\mathfrak{K}(W_{0})}  &  =H_{1},\text{
}\mathfrak{K}(W_{0})\cap i\mathfrak{K}(W_{0})=0\nonumber
\end{align}

It is important to be aware that, unlike QM, we are here dealing with real
(closed) subspaces $\mathfrak{K}$ of the complex one-particle Wigner
representation space $H_{1}$. An alternative which avoids the use of real
subspaces is to directly deal with complex dense subspaces $H_{1}%
(W_{0})=\mathfrak{K}(W_{0})+i\mathfrak{K}(W_{0})$ as in the third line.
Introducing the graph norm of the dense space the complex subspace in the
third line becomes a Hilbert space in its own right. The second and third line
require some explanation. The upper dash on regions denotes the causal
disjoint (which is the opposite wedge) whereas the dash on real subspaces
means the symplectic complement with respect to the symplectic form
$Im(\cdot,\cdot)$ on $H_{1}.$

The two properties in the third line are the defining relations of what is
called the \textit{standardness property} of a real
subspace\footnote{According to the Reeh-Schlieder theorem a local algebra
$\mathcal{A(O})$ in QFT is in standard position with respect to the vacuum
i.e. it acts on the vacuum in a cyclic and separating manner. The spatial
standardness, which follows directly from Wigner representation theory, is
just the one-particle projection of the Reeh-Schlieder property.}; any
standard K space permits to define an abstract s-operator%
\begin{align}
\mathfrak{s}(\psi+i\varphi) &  =\psi-i\varphi\label{inv}\\
\mathfrak{s} &  =\mathfrak{j}\delta^{\frac{1}{2}}\nonumber
\end{align}
whose polar decomposition (written in the second line) yields two modular
objects, a unitary modular group $\delta^{it}$ and a antiunitary reflection
which generally have however no geometric significance. The domain of the
Tomita $\mathfrak{s}$-operator is the same as the domain of $\delta^{\frac
{1}{2}}$ namely the real sum of the K space and its imaginary multiple. Note
that this domain is determined solely in terms of Wigner group representation theory.

It is easy to obtain a net of K-spaces by $U(a,\Lambda)$-transforming the
K-space for the distinguished $W_{0}.$ A bit more tricky is the construction
of sharper localized subspaces via intersections
\begin{equation}
\mathfrak{K}(\mathcal{O})=%
{\displaystyle\bigcap\limits_{W\supset\mathcal{O}}}
\mathfrak{K}(W)
\end{equation}
where $\mathcal{O}$ denotes a causally complete smaller region (noncompact
spacelike cone, compact double cone). Intersection may not be standard, in
fact they may be zero in which case the theory allows localization in $W$ (it
always does) but not in $\mathcal{O}.$ Such a theory is still causal but not
local in the sense that its associated free fields are pointlike. One can show
that the intersection for spacelike cones $\mathcal{O=C}$ for all positive
energy is always standard. A standard subspace is uniqely affiliated with a
Tomita s-involution (\ref{inv}).

At this point the important question arises why, if these localization
subspaces are important for particle physics they did not appear already at
the time of Wigner?  After all, unlike the Wigner position operators, these
spaces are frame independent (covariantly defined) and for two causally
separated regions $\mathcal{O}_{1}$ and $\mathcal{O}_{2}$ regions the
simplectic inner product vanishes%
\begin{align}
& Im(\psi_{1},\psi_{2})=0,~\psi_{i}\in H(\mathcal{O}_{i})\\
& \left[  \Phi(\psi_{1}),\Phi(\psi_{2})\right]  =0\nonumber
\end{align}
Hence the symplectic inner product of modular localized one-particle wave
functions is nothing else than the free field commutator function: the modular
localization preempts the algebraic structure of free fields without having
the use of any quantization formalism. Naturally this would have been of great
interest to Wigner, but the modular localization concepts were only available
more than half a century later. 

Note that the relativistic DPI setting also starts from Wigner particles but
it completely ignores the presence of this modular localization structure
which, would anyhow not be consistent with the DPI interactions. 

There are three classes of irreducible positive energy representation, the
family of massive representations $(m>0,s)$ with half-integer spin $s$ and the
family of massless representation which consists really of two subfamilies
with quite different properties namely the $(0,h=~$half-integer$)$ class,
often called the neutrino-photon class, and the rather large class of
$(0,\kappa>0)$ infinite helicity representations parametrized by a
continuous-valued Casimir invariant $\kappa$ \cite{MSY}$.$

For the first two classes the $\mathfrak{K}$-space the standardness property
also holds for double cone intersections $\mathcal{O=D}$ for arbitrarily small
$\mathcal{D},$ but this is definitely not the case for the infinite helicity
family for which the localization spaces for compact spacetime regions turn
out to be trivial\footnote{It is quite easy to prove the standardness for
spacelike cone localization (leading to singular stringlike generating fields)
just from the positive energy property which is shared by all three families
\cite{BGL}.}. Passing from localized subspaces $\mathcal{K}$ in the
representation theoretical setting to singular covariant generating wave
functions (the first quantized analogs of generating fields) one can show that
the $\mathcal{D}$ localization leads to pointlike singular generators
(state-valued distributions) whereas the spacelike cone localization
$\mathcal{C}$ is associated with semiinfinite spacelike stringlike singular
generators \cite{MSY}. Their second quantized counterparts are pointlike or
stringlike covariant fields. It is remarkable that one does not need to
introduce generators which are localized on hypersurfaces (branes).

Although the observation that the third Wigner representation class is not
pointlike generated was made many decades ago, the statement that it is
semiinfinite string-generated and that this is the worst possible case of
state localization (which needs the knowledge of modular theory) is of a more
recent vintage \cite{BGL}\cite{MSY}.

But what is the physical significance of modular localization of wave function
which, different from the probabilistic BNW localized states are obviously
frame-independent and hence cannot be used for describing the dissipation of
wave packets and the related scattering theory? The answer is that they are
the projections of the dense subspaces (the Reeh-Schlieder domains) generated
by applying modular extension of the localized subalgebra $\mathcal{A}%
(\mathcal{O})$ to the vacuum onto the one-particle space%
\begin{align}
& P_{1}H(\mathcal{O})=\mathfrak{K}(\mathcal{O})+i\mathfrak{K}(\mathcal{O}%
)\equiv H_{1}(\mathcal{O})\\
H(\mathcal{O})  & =domS(\mathcal{O}),\text{ }S(\mathcal{O})A\Omega=A^{\ast
}\Omega,~A\in\mathcal{A(O})\nonumber
\end{align}

In other words the one-particle dense localization spaces are projections of
the Reeh-Schlieder spaces\footnote{At the time of the discovery of the density
of the spaces $\mathcal{A(O})\Omega$ the modular theory was not yet known.
There is no change of content if one uses the same terminology for their
modular extension $domS(\mathcal{O}).$}.  

There is a very subtle aspect of modular localization which one encounters in
the second Wigner representation class of massless finite helicity
representations (the photon, graviton..class). Whereas in the massive case all
spinorial fields $\Psi^{(A,\dot{B})}$ the relation of the physical spin $s$
with the two spinorial indices follows the naive angular momentum composition
rules \cite{Weinbook}%
\begin{align}
\left\vert A-\dot{B}\right\vert  &  \leq s\leq\left\vert A+\dot{B}\right\vert
,\text{ }m>0\label{line}\\
s  &  =\left\vert A-\dot{B}\right\vert ,~m=0\nonumber
\end{align}
the second line contains the significantly reduced number of spinorial
descriptions for zero mass and finite helicity representations. What is going
on here, why is there, in contradistinction to classical field theory no
covariant s=1 vector-potential $A_{\mu}$ or no $g_{\mu\nu}$ in case of s=2 ?
Why are the admissible covariant generators of the Wigner representation in
this case limited to field strengths (for s=2 the linearized Riemann tensor)?

The short answer is that all these missing generators exist as stringlike
covariant objects, the above restriction in the massless case only results
from the covariantization to pointlike generators. The full range of spinorial
possibilities (\ref{line}) returns in terms of string localized fields
$\Psi^{(A,\dot{B})}(x,e)$ if $s\neq\left\vert A-\dot{B}\right\vert $. These
generating free fields are covariant and "string-local"%

\begin{align}
U(\Lambda)\Psi^{(A,\dot{B})}(x,e)U^{\ast}(\Lambda)  &  =D^{(A,\dot{B}%
)}(\Lambda^{-1})\Psi^{(A,\dot{B})}(\Lambda x,\Lambda e)\label{string}\\
\left[  \Psi^{(A,\dot{B})}(x,e),\Psi^{(A^{\prime},\dot{B}^{\prime})}%
(x^{\prime},e^{\prime}\right]  _{\pm}  &  =0,~x+\mathbb{R}_{+}e><x^{\prime
}+\mathbb{R}_{+}e^{\prime}\nonumber
\end{align}
Here the unit vector $e$ is the spacelike direction of the semiinfinite string
and the last line expresses the spacelike fermionic/bosonic spacelike
commutation. The best known illustration is the ($m=0,s=1$) vectorpotential
representation; in this case it is well-known that although a generating
pointlike field strength exists, there is no \textit{pointlike}
vectorpotential acting in a Hilbert space.

According to (\ref{string}) the modular localization approach offers as a
substitute a stringlike covariant vector potential $A_{\mu}(x,e).$ In the case
($m=0,s=2$) the "field strength" is a fourth degree tensor which has the
symmetry properties of the Riemann tensor (it is often referred to as the
\textit{linearized} Riemann tensor). In this case the string-localized
potential is of the form $g_{\mu\nu}(x,e)$ i.e. resembles the metric tensor of
general relativity. Some consequences of this localization for a reformulation
of gauge theory will be mentioned in section 8.

Even in case of massive free theories where the representation theoretical
approach of Wigner does not require to go beyond pointlike localization,
covariant stringlike localized fields exist. Their attractive property is that
they improve the short distance behavior e.g. a massive pointlike
vector-potential of \textit{sdd=2} passes to a string localized vector
potential of \textit{sdd=1}. In this way the increase of the sdd of pointlike
fields with spin s can be traded against string localized fields of spin
independent dimension with sdd=1. This observation would suggest the
possibility of an enormous potential enlargement of perturbatively accessible
higher spin interaction in the sense of power counting.

A different kind of spacelike string-localization arises in d=1+2 Wigner
representations with anomalous spin \cite{Mu1}. The amazing power of the
modular localization approach is that it preempts the spin-statistics
connection already in the one-particle setting, namely if s is the spin of the
particle (which in d=1+2 may take on any real value) then one finds for the
connection of the symplectic complement with the causal complement the
generalized duality relation
\begin{equation}
\mathfrak{K}(\mathcal{O}^{\prime})=Z\mathfrak{K}(\mathcal{O})^{\prime}%
\end{equation}
where the square of the twist operator $Z=e^{\pi is}~$is easily seen (by the
connection of Wigner representation theory with the two-point function) to
lead to the statistics phase $=Z^{2}$ \cite{Mu1}.

The fact that one never has to go beyond string localization (and fact, apart
from $s\geq1$, never beyond point localization) in order to obtain generating
fields for a QFT is remarkable in view of the many attempts to introduce
extended objects into QFT.

It is helpful to be again reminded that modular localization which goes with
real subspaces (or dense complex subspaces), unlike BNW localization, cannot
be connected with probabilities and projectors. It is rather related to causal
localization aspects; the standardness of the K-space for a compact region is
nothing else then the one-particle version of the Reeh-Schlieder property. As
will be seen in the next section modular localization is also an important
tool in the non-perturbative construction of interacting models.

\section{Algebraic aspects of modular theory}

A net of real subspaces $\mathfrak{K}(\mathcal{O})$ $\subset$ $H_{1}$ for an
finite spin (helicity) Wigner representation can be "second
quantized"\footnote{The terminology 2$^{nd}$ quantization is a misdemeanor
since one is dealing with a rigorously defined functor within QT which has
little in common with the artful use of that parallellism to classical theory
called "quantization". In Edward Nelson's words: (first) quantization is a
mystery, but second quantization is a functor.} via the CCR (Weyl)
respectively CAR quantization functor; in this way one obtains a covariant
$\mathcal{O}$-indexed net of von Neumann algebras $\mathcal{A(O)}$ acting on
the bosonic or fermionic Fock space $H=Fock(H_{1})$ built over the
one-particle Wigner space $H_{1}.$ For integer spin/helicity values the
modular localization in Wigner space implies the identification of the
symplectic complement with the geometric complement in the sense of
relativistic causality, i.e. $\mathfrak{K}(\mathcal{O})^{\prime}%
=\mathfrak{K}(\mathcal{O}^{\prime})$ (spatial Haag duality in $H_{1}$). The
Weyl functor takes this spatial version of Haag duality into its algebraic
counterpart. One proceeds as follows: for each Wigner wave function
$\varphi\in H_{1}$ the associated (unitary) Weyl operator is defined as%
\begin{align}
Weyl(\varphi)  &  :=expi\{a^{\ast}(\varphi)+a(\varphi)\}\in B(H)\\
\mathcal{A(O})  &  :=alg\{Weyl(\varphi)|\varphi\in\mathfrak{K}(\mathcal{O}%
)\}^{^{\prime\prime}},~~\mathcal{A(O})^{\prime}=\mathcal{A(O}^{\prime
})\nonumber
\end{align}
where $a^{\ast}(\varphi)$ and $a(\varphi)$ are the usual Fock space creation
and annihilation operators of a Wigner particle in the wave function $\varphi
$. We then define the von Neumann algebra corresponding to the localization
region $\mathcal{O}$ in terms of the operator algebra generated by the
functorial image of the modular constructed localized subspace $\mathfrak{K}%
(\mathcal{O})$ as in the second line. By the von Neumann double commutant
theorem, our generated operator algebra is weakly closed by definition.

The functorial relation between real subspaces and von Neumann algebras via
the Weyl functor preserves the causal localization structure and hence the
spatial duality passes to its algebraic counterpart. The functor also commutes
with the improvement of localization through intersections $\cap$ according to
$\mathfrak{K}(\mathcal{O})=\cap_{W\supset O}\mathfrak{K}(W),~\mathcal{A(O}%
)=\cap_{W\supset O}\mathcal{A}(W)$ as expressed in the commuting diagram%
\begin{align}
&  \left\{  \mathfrak{K}(W)\right\}  _{W}\longrightarrow\left\{
\mathcal{A}(W)\right\}  _{W}\\
&  \ \ \downarrow\cap~~~\ \ \ \ \ \ \ \ \ \ ~\ ~\downarrow\cap\nonumber\\
~~ &  \ \ \ \mathfrak{K}(\mathcal{O})\ \ \ \longrightarrow\ \ ~\mathcal{A(O}%
)\nonumber
\end{align}
Here the vertical arrows denote the tightening of localization by intersection
whereas the horizontal ones denote the action of the Weyl functor. This
commuting diagram expresses the functorial relation between particles and
fields in the absence of interactions. In the interacting case the loss of the
diagram and the unsolved particle-field problems are synonymous. It is also
the reason why, in contrast to QM, the existence problem of interacting QFTs
even after more than 80 years remains unsolved. The wedge regions continue to
play a distinguished role in attempts to construct interacting models (for
some modular successes in d=1+1 see below).  

The case of half-integer spin representations is analogous \cite{Fa-Sc}, apart
from the fact that there is a mismatch between the causal and symplectic
complements which must be taken care of by a \textit{twist operator}
$\mathcal{Z}$ and as a result one has to use the CAR functor instead of the
Weyl functor.

In case of the large family of irreducible zero mass infinite spin
representations in which the lightlike little group is faithfully represented,
the finitely localized K-spaces are trivial $\mathfrak{K}(\mathcal{O})=\{0\}$
and the \textit{most tightly localized nontrivial spaces} \textit{are of the
form} $\mathfrak{K}(\mathcal{C})$ for $\mathcal{C}$ an arbitrarily narrow
\textit{spacelike cone}. As a double cone contracts to its core which is a
point, the core of a spacelike cone is a \textit{covariant spacelike
semiinfinite string}. The above functorial construction works the same way for
the Wigner infinite spin representation, except that in that case there are no
nontrivial algebras which have a smaller localization than $\mathcal{A(C})$
and there is no field which is sharper localized than a semiinfinite string.
As stated before, stringlike generators, which are also available in the
pointlike case, turn out to have an improved short distance behavior which
makes them preferable from the point of view of formulating interactions
within the power counting limit. They can be constructed from the unique
Wigner representation by so called intertwiners between the unique canonical
and the many possible covariant (dotted-undotted spinorial) representations.
The Euler-Lagrange aspects plays no direct role in these construction since
the causal aspect of hyperbolic differential propagation are fully taken care
of by modular localization and also because most of the spinorial higher spin
representations (\ref{line}) anyhow cannot be characterized in terms of
Euler-Lagrange equations. The modular localization is the more general method
of implementating causal propagation than that from hyperbolic equations of motions.

A basis of local covariant field coordinatizations is then defined by Wick
composites of the free fields. The case which deviates furthest from classical
behavior is the pure stringlike infinite spin case which relates a
\textit{continuous} family of free fields with one irreducible infinite spin
representation. Its non-classical aspects, in particular the absence of a
Lagrangian, is the reason why the spacetime description in terms of
semiinfinite string fields has been discovered only recently rather than at
the time of Jordan's field quantization or Wigner's representation theoretical approach.

Using the standard notation $\Gamma$ for the second quantization functor which
maps real localized (one-particle) subspaces into localized von Neumann
algebras and extending this functor in a natural way to include the images of
the $\mathfrak{K}(\mathcal{O})$-associated $s,\delta,j$ which are denoted by
$S,\Delta,J,$ one arrives at the Tomita Takesaki theory of the
interaction-free local algebra ($\mathcal{A(O}),\Omega$) in standard
position\footnote{The functor $\Gamma$ preserves the standardness i.e. maps
the spatial one-particle standardness into its algebraic counterpart.}%
\begin{align}
&  H_{Fock}=\Gamma(H_{1})=e^{H_{1}},~\left(  e^{h},e^{k}\right)
=e^{(h,k)}\label{mod}\\
&  \Delta=\Gamma(\delta),~J=\Gamma(j),~S=\Gamma(s)\nonumber\\
&  SA\Omega=A^{\ast}\Omega,~A\in\mathcal{A}(O),~S=J\Delta^{\frac{1}{2}%
}\nonumber
\end{align}

With this we arrive at the core statement of the Tomita-Takesaki theorem which
is a statement about the two modular objects $\Delta^{it}$ and $J$ on the
algebra%
\begin{align}
\sigma_{t}(\mathcal{A(O}))  &  \equiv\Delta^{it}\mathcal{A(O})\Delta
^{-it}=\mathcal{A(O})\\
J\mathcal{A(O})J  &  =\mathcal{A(O})^{\prime}=\mathcal{A(O}^{\prime})\nonumber
\end{align}
in words: the reflection $J$ maps an algebra (in standard position) into its
von Neumann commutant and the unitary group $\Delta^{it}$ defines an
one-parametric automorphism-group $\sigma_{t}$ of the algebra. In this form
(but without the last geometric statement involving the geometrical causal
complement $\mathcal{O}^{\prime})$ the theorem hold in complete mathematical
generality for standard pairs ($\mathcal{A},\Omega$). The free fields and
their Wick composites are "coordinatizing" singular generators of this
$\mathcal{O}$-indexed net of operator algebras in the sense that the smeared
fields $A(f)$ with $suppf\subset\mathcal{O}$ are (unbounded operators)
affiliated with $\mathcal{A(O})$ and in a certain sense generate
$\mathcal{A(O}).$

In the above second quantization context the origin of the T-T theorem and its
proof is clear: the symplectic disjoint passes via the functorial operation to
the operator algebra commutant (\ref{line}) and the spatial one-particle
automorphism goes into its algebraic counterpart. The definition of the Tomita
involution $S$ through its action on the dense set of states (guarantied by
the standardness of $\mathcal{A}$) as $SA\Omega=A^{\ast}\Omega$ and the action
of the two modular objects $\Delta,J$ (\ref{mod}) is part of the general
setting of the modular Tomita-Takesaki theory of abstract operator algebras in
"standard position"; standardness is the mathematical terminology for the
physicists Reeh-Schlieder property i.e. the existence\footnote{In QFT any
finite energy vector (which of course includes the vacuum) has this property
as well as any nondegenerated KMS state. In the mathematical setting it is
shown that standard vectors are "$\delta-$dense" in $H$.} of a vector
$\Omega\in H$ with respect to which the algebra acts cyclic and has no
"annihilators" of $\Omega.$ Naturally the proof of the abstract T-T theorem in
the general setting of operator algebras is more involved\footnote{The local
algebras of QFT are (as a consequence of the split property) hyperfinite; for
such operator algebras Longo has given an elegant proof \cite{hyper}.}.

The domain of the unbounded Tomita involution $S$ turns out to be
"kinematical" in the sense that the dense set which features in the
Reeh-Schlieder theorem is determined in terms of the representation of the
connected part of the Poincar\'{e} group i.e. the particle/spin
spectrum\footnote{For a wedge $W$ the domain of $S_{W}$ is determined in terms
of the domain of the "analytic continuation" $\Delta_{W}^{\frac{1}{2}}$ of the
wedge-associated Lorentz-boost subgroup $\Lambda_{W}(\chi),$ and for subwedge
localization regions $\mathcal{O}$ the dense domain is obtained in terms of
intersections of wedge domains.}. In other words the Reeh-Schlieder domains in
an interacting theory with asymptotic completeness are identical to those of
the incoming or outgoing free field theory.

The important property which renders this useful beyond free fields as a new
constructive tool in the presence of interactions, is that for $\left(
\mathcal{A}(W),\Omega\right)  ~$ the antiunitary involution $J$ depends on the
interaction, whereas $\Delta^{it}$ continues to be uniquely fixed by the
representation of the Poincar\'{e} group i.e. by the particle content. In fact
it has been known for some \cite{Sch} time that $J$ is related with its free
counterpart $J_{0}$ through the scattering matrix%
\begin{equation}
J=J_{0}S_{scat} \label{scat}%
\end{equation}

This modular role of the scattering matrix as a relative modular invariant
between an interacting theory and its free counterpart comes as a surprise. It
is precisely this property which opens the way for an inverse scattering
construction. If one only looks at the dense localization of states which
features in the Reeh-Schlieder theorem, one misses the dynamics. There is
presently no other way to inject dynamics than generating these states by
applying operators from operator algebras. The properties of $J$ are
essentially determined by the relation of localized operators $A~$to their
Hermitian adjoints $A^{\ast}$\footnote{According to a theorem of Alain Connes
\cite{Connes} the existence of operator algebras in standard position can be
inferred if the real subspace $K$ permit a decompositions into a natural
positive cone and its opposite with certain facial properties of positive
subcones. Although this construction has been highly useful in Connes
classification of von Neumann factors, it has not yet been possible to relate
this to physical concepts.}.

The physically relevant facts emerging from modular theory can be condensed
into the following statements:

\begin{itemize}
\item \textit{The domain of the unbounded operators }$S(\mathcal{O})$\textit{
is fixed in terms of intersections of the wedge domains associated to
}$S(W);~$\textit{in other words it is determined by the particle content alone
and therefore of a kinematical nature. These dense domains change with
}$\mathcal{O}$\textit{ i.e. the dense set of localized states has a bundle
structure.}

\item \textit{The complex domains }$DomS(\mathcal{O})=K(\mathcal{O}%
)+iK(\mathcal{O})$\textit{ decompose into real subspaces }$K(\mathcal{O}%
)=\overline{\mathcal{A(O})^{sa}\Omega}.$\textit{ This decomposition contains
dynamical information which in case }$\mathcal{O}=W$\textit{ reduces to the
S-matrix (\ref{scat}). Assuming the validity of the crossing properties for
formfactors, the S-matrix fixes }$\mathcal{A}(W)$\textit{ uniquely \cite{S3}.}
\end{itemize}

The remainder of this subsection contains some comments about a remarkable
constructive success of these modular methods with respect to a particular
family of interacting theories. For this one needs some additional
terminology. Let us enlarge the algebraic setting by admitting unbounded
operators with Wightman domains which are affiliated to $\mathcal{A(O})$ and
let us agree to just talk about "$\mathcal{O}$-localized operators" when we do
not want to distinguish between bounded and affiliated unbounded operators. We
call an $\mathcal{O}$-localized operators a vacuum \textbf{p}olarization
\textbf{f}ree \textbf{g}enerator (PFG) if applied to the vacuum it generated a
one-particle state without admixture of a vacuum-polarization cloud. The
following three theorems have turned out to be useful in a constructive
approach based on modular theory.

\textbf{Theorem} (\cite{BBS}): \textit{The existence of an }$\mathcal{O}%
$\textit{-localized PFG for a causally complete subwedge region }%
$\mathcal{O}\subset W$\textit{ implies the absence of interactions i.e. the
generating fields are (}a slight generalization \cite{BBS} of the Jost-Schroer
theorem (referred to in \cite{Wigh}\cite{St}) which still used the existence
of pointlike covariant fields).

\textbf{Theorem (}\cite{BBS}\textbf{)}: \textit{Modular theory for wedge
algebras insures the existence of wedge-localized PFGs. Hence the wedge region
permits the best compromise between interacting fields and one-particle
states}\footnote{It is the smallest causally closed region (its localization
representing a field aspect) which contains one-particle creators.}\textit{. }

\textbf{Theorem (}\cite{BBS}\textbf{)}: \textit{Wedge localized PFGs with good
(Wightman-like) domain properties ("temperate" PFGs) lead to the absence of
particle creation (pure elasic }$S_{scat})$\textit{ which in turn is only
possible in d=1+1 and leads to the factorizing models (which hitherto were
studied in the setting of the bootstrap-formfactor program \cite{Ba-Ka}). The
compact localized interacting subalgebras }$\mathcal{A(O})$\textit{ have no
PFGs and possess the full interaction-induced vacuum polarization clouds.}

Some additional comments will be helpful. The first theorem gives an intrinsic
(not dependent on any Lagrangian or other extraneous properties) local
definition of the presence of interaction, even though it is not capable to
differentiate between different kind of interactions (which would be reflected
in the shapes of interaction-induced polarization clouds). The other two
theorems suggest that the knowledge of the wedge algebra $\mathcal{A(}%
W)\subset B(H)$ may serve as a useful starting point for classifying and
constructing models of LQP in a completely intrinsic fashion. Knowing
generating operators of $\mathcal{A(}W)$ including their transformation
properties under the Poincar\'{e} group is certainly sufficient and
constitutes the most practical way for getting the construction started (for
additional informations see later section).

All wedge algebras possess affiliated PFGs but only in case they come with
reasonable domain properties ("temperate") they can presently be used in
computations. This requirement only leaves models in d=1+1 which in addition
must be factorizing (integrable); in fact the modular theory used in
establishing these connections shows that there is a deep connection between
integrability in QFT and vacuum polarization properties \cite{BBS}.

Temperate PFGs which generate wedge algebra for factorizing models have a
rather simple algebraic structure. They are of the form (in the absence of
boundstates)%
\begin{equation}
Z(x)=\int\left(  \tilde{Z}(\theta)e^{-ipx}+h.c.\right)  \frac{dp}{2p_{0}}
\label{Z}%
\end{equation}
where in the simplest case $\tilde{Z}(\theta),\tilde{Z}^{\ast}(\theta)~$are
one-component objects\footnote{This case leads to the Sinh-Gordon theory and
related models.} which obey the Zamolodchikov-Faddeev commutation relations
\cite{S3}. In this way the formal Z-F device which encoded the two-particle
S-matrix into the commutation structure of the Z-F algebra receives a profound
spacetime interpretation. Like free fields these wedge fields are on mass
shell, but their Z-F commutation relations renders them non-local, more
precisely wedge-local \cite{S3}.

The simplicity of the wedge generators in factorizing models is in stark
contrast to the richness of compactly localized operators e.g. of operators
affiliated to a spacetime double cone $\mathcal{D}$ which arises as a relative
commutant $\mathcal{A(D})=\mathcal{A}(W_{a})^{\prime}\cap\mathcal{A}(W)$. The
wedge algebra $\mathcal{A}(W)$ has simple generators and the full space of
formal operators affiliated with $\mathcal{A(}W)$ has the form of an infinite
series in the Z-F operators with coefficient functions $a(\theta_{1}%
,...\theta_{n})$ with analyticity properties in a $\theta$-strip%
\begin{equation}
A(x)=%
{\displaystyle\sum}
\frac{1}{n!}%
{\displaystyle\int_{\partial S(0,\pi)}}
d\theta_{1}...%
{\displaystyle\int_{\partial S(0,\pi)}}
d\theta_{n}e^{-ix\sum p(\theta_{i})}a(\theta_{1},...\theta_{n}):\tilde
{Z}(\theta_{1})...\tilde{Z}(\theta_{1}): \label{Zam}%
\end{equation}
where for the purpose of a compact notation we view the creation part
$\tilde{Z}^{\ast}(\theta)$ as $\tilde{Z}(\theta+i\pi)$ i.e. as coming from the
upper part of the strip $S(0,\pi)$\footnote{The notation is suggested by the
the strip analyticity coming from wedge localization. Of course only certain
matrix elements and expectation values, but not field operators or their
Fourier transforms, can be analytic; therefore the notation is symbolic.}. The
requirement that the series (\ref{Zam}) commutes with the translated generator
$A(f_{a})\equiv U(a)A(f)U^{\ast}(a)$ affiliated with $\mathcal{A}(W_{a})$
defines formally a subspace of operators affiliated with $\mathcal{A(D}%
)=\mathcal{A}(W_{a})^{\prime}\cap\mathcal{A}(W).$

As a result of the simplicity of the $\tilde{Z}$ generators one can
characterize these subspaces in terms of analytic properties of the
coefficient functions $a(\theta_{1},...\theta_{n}).$ The latter are related to
the formfactors of $A$ which are the matrix elements of $A$ between "ket" in
and "bra" out particle states. The coefficient functions in (\ref{Zam}) obey
the crossing property. In this way the computational rules of the
bootstrap-formfactor program \cite{Ba-Ka} are explained in terms of an
algebraic construction \cite{Sch}.

This is similar to the old Glaser-Lehmann-Zimmermann representation for the
interacting Heisenberg field \cite{GLZ} in terms of incoming free field. Their
use has the disadvantage that the coefficient functions are not related by the
crossing property to one analytic master function. The convergence of both
series has remaind an open problem. So unlike the perturbative series
resulting from renormalized perturbation theory which have been shown to
diverge even in models with optimal short distance behavior (even Borel
resummability does not help), the status of the GLZ and formfactor series
remains unresolved.

The main property one has to establish, if one's aim is to secure the
existence of a QFT with local observables, is the standardness of the double
cone intersection $\mathcal{A(D})=\cap_{W\supset\mathcal{D}}\mathcal{A(}W).$
Based on nuclearity properties of degrees of freedom in phase space
(discovered by Buchholz and Wichmann \cite{Bu}), Lechner has established the
standardness of these intersections and in this way demonstrated the
nontriviality of the model as a localized QFT \cite{Lech1}\cite{Lech2}. For
the first time in the history of QFT one now has a construction method which
goes beyond the Hamiltonian- and measure-theoretical approach of the 60s
\cite{G-J}. The old approach could only deal with superrenormalizable models
i.e. models whose basic fields did not have a short distance dimension beyond
that of a free field.

The factorizing models form an interesting theoretical laboratory where
problems, which accompanied QFT almost since its birth, resurface in a
completely new light. The very existence of these theories, whose fields have
anomalous trans-canonical short distance dimensions with interaction-dependent
strengths, shows that there is nothing intrinsically threatening about
singular short distance behavior. Whereas in renormalized perturbation theory
the power counting rule only permits logarithmic corrections to the canonical
(free field) dimensions, the construction of factorizing models starting from
wedge algebras and their $Z$ generators allow arbitrary high powers. That many
problems of QFT are not intrinsic but rather caused by a particular method of
quantization had already been suspected by the protagonist of QFT Pascual
Jordan who, as far back as 1929, pleaded for a formulation "without (classic)
crutches" \cite{Jo}. The above construction of factorizing models which does
not use any of the quantization schemes and in which the model does not even
come with a Lagrangian name may be considered at the first realization of
Jordan's plea at which he arrived on purely philosophically grounds.

The significant conceptual distance between QM and LQP begs the question in
what sense the statement that QM is a nonrelativistic limit of LQP should be
understood. By this we do not mean a formal manipulation in a Lagrangian or
functional integral representation, but an argument which starts from the
correlation functions or operator algebras of an interacting LQP and explains
in what way an interacting QFT looses its modular localization + vacuum
polarization and moves into the conceptual setting of QM. This is far from
evident since in certain cases as that of 3-dimensional plektonic statistics
the nonrelativistic limit retains the vacuum polarization, which is necessary
to sustain the braid group statistics and thus becomes a nonrelativistic QFT
instead of QM.

Apparently such arguments do not yet exist. One attempt in this direction
could consist in starting from the known formfactors of a factorizing model
(as e.g. the Sinh-Gordon model) and study the simplifications for small
rapidity $\theta.$ An insight of this kind would constitute an essential
improvement of our understanding of the QM-QFT interface.

Since modular theory continues to play an important role in the remaining
section as well as part II, some care is required in avoiding potential
misunderstandings. It is crucial to be aware of the fact that by restricting
the global vacuum state to, a say double cone algebra $\mathcal{A(D})$
whereupon it becomes a thermal KMS state, there is no change in the values of
the global vacuum expectation values%
\begin{equation}
(\Omega_{vac},A\Omega_{vac})=\left(  \Omega_{\operatorname{mod},\beta}%
,A\Omega_{\operatorname{mod},\beta}\right)  ,~\text{ }A\in\mathcal{A(D})
\end{equation}
where for the standard normalization of the modular Hamiltonian\footnote{The
modular Hamiltonian lead to fuzzy motions within $\mathcal{A(O})$ except in
case of $\mathcal{O}=W$ when the modular Hamiltonian is identical to the boost
generator.} $\beta=-1.$ This notation on the right hand side means that the
vacuum expectation values, if restricted to $A\in\mathcal{A(D}),$ fulfill an
additional property (which without the restriction to the local algebra would
not hold), namely the KMS relation%
\begin{equation}
\left(  \Omega_{\operatorname{mod},\beta},AB\Omega_{\operatorname{mod},\beta
}\right)  =\left(  \Omega_{\operatorname{mod},\beta},B\Delta_{\mathcal{A(O}%
)}A\Omega_{\operatorname{mod},\beta}\right)  \label{ther}%
\end{equation}
At this point one may wonder how a global vacuum state can turn into a thermal
state on a smaller algebra without any thermal exchange taking place. The
answer is that the in terms of ($\mathcal{A}(\mathcal{D}),\Omega_{vac}$)
canonically defined modular Hamiltonian $K_{\operatorname{mod}}$ with
$\Delta=e^{-K_{\operatorname{mod}}}$ is very different from the original
translative Hamiltonian $H_{tr}$ whose lowest energy eigenstate defines the
vacuum, whereas $K_{\operatorname{mod}}$ is the generator of a modular
automorphism of $\mathcal{A}(\mathcal{D})$ which in the geometric terminology
preferred by physicists (even when it becomes inappropriate) describes a
"fuzzy" motion inside $\mathcal{D}.$

The modular automorphism is actually defined on the global algebra $B(H)$
where it acts in such a way that $\mathcal{A}(\mathcal{D})$ and $\mathcal{A}%
(\mathcal{D})^{\prime}=\mathcal{A}(\mathcal{D}^{\prime})$ are automorphically
mapped into themselves. The state vector $\Omega_{vac}\in H$ is a zero
eigenvalue of $K_{\operatorname{mod}}$ which sits in the middle of a symmetric
two-sided spectrum. What has changed through the process of restriction is not
the state but rather the way of looking at it: $H_{\operatorname{mod}}$
describes the dynamics of an "observer" confined to $\mathcal{D}$ whereas
$H_{tr}$ has obviously no intrinsic meaning in a world restricted to
$\mathcal{D}.$ In fact it turns out that the fuzzy automorphism becomes
geometric near the causal horizon of the region $\mathcal{O}$ (see second part)

The thermal aspect of modular theory refers to the modular Hamiltonian; it
does not mean that one is creating heat with respect to the usual inertial
frame Hamiltonian; its energy conservation is always maintained and
observer-relevant heat is never generated as long as the observer's system
remains inertial. Already in this context of inertial observer in the ground
state and a modular observer for whom this state becomes thermal, the
attentive reader may correctly presume an anticipation of the thermal
manifestations of black holes as localized restrictions of a larger system
(the Kruskal extension of the Schwartzschild black hole).

Going back to the Unruh Gedankenexperiment featuring a non-inertial observer
which in order to follow the path of the modular Hamiltonian of a Rindler
wedge W must be uniformely accelerated in some spatial direction, the standard
question is the thermal aspect of the W-reduced vacuum real or is it a
mathematical aspect carried too far ? The Unruh effect claims that this is
really what the non-inertial observer measures in his taken along counter.
Although the effect is so tiny that it will probably never be observed, the
existence of the thermal radiation is a inescapable consequence of our most
successful theories. One is accustomed to all kind of forces in noninertial
systems but where does the nonzero thermal radiation density come from?

In order to create a causal horizon the observer must be uniformely
accelerated which requires feeding energy into the system. In other words the
realization of the innocent looking restriction in localization requires an
enormous energy expenditure thus revealing in one example what is behind the
physics of the harmless sounding word "restriction". Only when the modular
Hamiltonian describes a movement which corresponds to a diffeomorphism of
spacetime is there a chance to think in terms of an Unruh kind of
Gedankenexperiment. As was explained before the modular situation is more
physical in black hole situations where the position of event horizons is
fixed by the metric independent of what an observer does. This is underlined
by the earlier mentioned existence of a pure state on the Kruskal extension of
the Schwarzschild solution (the Hartle-Hawking state); this state has the
position of the event horizon worked in and does not need any observer for its
definition. Restricted to the region outside of the black hole the modular
automorphism describes the timelike Killing movement which is as close as one
can come to an inertial path. The correponding Killing Hamiltonian is the
closest analog of the inertial Hamiltonian in Minkowski spacetime.

There remains the question to what extent quantum physics in an Unruh frame is
different from that in an inertial frame. There are no particles (in the sense
of Wigner) since the vacuum behaves like a thermal densitiy in which counter
experiments only permit the measurement of radiation densities as in standard
thermal radiation or cosmic microwave background radiation. In fact it is
quite straightforward to show the \textit{LSZ scattering limit does not exist
in the Unruh boost time}, a fact which is related to the two-sided spectrum of
the modular Hamiltonian with respect to the W-reduced reduced ground state of
the original inertial system. To wit, the global zero temperature Wigner-Fock
space can be used also after the wedge restriction, but the global n-particle
states loose their intrinsic physical meaning. Apart from the modular aspects
the problems of the Unruh effect have been treated by many authors including
authors from the foundational community \cite{Cl-Hal}\cite{Arag}.

In fact there is a \textit{continuous family of modular "Hamiltonians"} which
are the generators the modular unitaries for sequences of included regions.
The modular Hamiltonian of the larger region will spread the smaller localized
algebra into the larger region.

Besides the thermal description of restricted states there is one other
macroscopic manifestation of vacuum polarization which has caused unbelieving
amazement in philosophical circles namely the cyclicity of the vacuum (the
Reeh-Schlieder property) with respect to algebras localized in arbitrarily
small spacetime region or in its more metaphoric presentation the idea that by
doing something in a small earthly laboratory for an arbitrary small fraction
of time one can approximate any state "behind the moon" with arbitrary
precision by (however with ever increasing expenditure in energy \cite{Haag}).

Both consequences of vacuum polarization and as such interconnected, they
\footnote{Sometimes used as a metaphor for the Reeh-Schlieder property.} are
manifestations of an \textit{holistic} behavior which in this extreme form is
absent in QM. Instead of the division between an object to be measured, the
measuring apparatus and the environment, without which the modern quantum
mechanical measurement theory cannot be formulated, in LQP such a separation
is called into question. By restricting the vacuum to the inside, one already
specifies the vacuum polarization driven dynamic on the causal disjoint. In
the "state behind the moon argument" the difficulty in a system-environment
dichotomy is even more palpable.

This is indeed an extremely surprising feature which goes considerably beyond
the kinematical change caused by entanglement as the result of the quantum
mechanical division into measured system and environment. It is this
dependence of the reduced vacuum state on the localization region inside which
it is tested with localized algebras which raises doubts about what are really
non-fleeting persistent properties of a material substance. The monad
description in the next section strengthens this little holistic aspect of LQP.

As we have seen the thermal aspects of modular localization are very rich from
an epistemic viewpoint. The ontological content of these observations on the
other hand is quite weak; it is only when the (imagined) causal localization
horizons passes from a Gedanken objects to a (real) event horizons through the
curvature of spacetime, that the fleeting aspect of causal horizons of
observers pass to an intrinsic ontological property of spacetime in the case
of black holes. But even if one's main interest is to do black hole physics,
it is wise to avoid a presently popular "shut up and compute" attitude and to
understand the conceptual basis in LQP of the thermal aspect of localization
and the peculiar thermal entanglement which contrasts the
information-theoretical quantum mechanical entanglement. Ignoring these
conceptual aspects one may easily be drawn into a fruitless and protractive
arguments as it happened (and still happens) with the entropy/information loss issue.

Up to now the terminology "localization" was used both for states and for
subalgebras. In the absence of interactions they are synonymous; this is
because free fields are uniquely determined by positive energy representations
of the Poincar\'{e}, in fact the generators of covariant wave functions pass
directly to generating fields. A representation which has no infinite spin
components is always pointlike generated. This applies in particular to string
theory which is a misnomer for infinite component field theory \cite{foun}.
Such a close relation between algebraic and state localization breaks down in
the presence of interactions. It is perfectly conceivable to have a theory
with "topological charges" \cite{Haag} which by definition cannot be described
by compactly localizable operators but need spacelike string localizable
generating fields. In that case the neutral observable algebra has the usual
compact localizability whereas the charge-carrying part of the total algebra
may need semiinfinite string generators for its description \cite{B-F}. The
fact that this possibility could even occur in massive QCD like theories makes
it very interesting, but unfortunately there is no illustrative example.

The problem of localization is of pivotal relevance for QFT. But nowhere is
glory and failure so interlinked as in this issue. The misunderstandings range
from the comparatively harmless confusion between the BNW localization of
states and the modular localization of observables to the very serious
misunderstanding of string theory. Besides these grave consequences the
innumerable confusions about particles, frame-independent localization of
states and algebras have been a harmless nuisance since they were comitted by
individuals and not by globalized communities. As the example of the
misinterpreted Fermi Gedankenexperiment showed, such mistakes can be corrected.

In particular the 10 dimensional covariant infinite component unitary
superstring representation of the Poincar\'{e} group coming from the
quantization of the bilinearized Nambu-Goto Lagrangian is according to the
before mentioned theorem (for representations which do not contain Wigner's
infinite spin representation) a pointlike localized object, and this also
applies to its predecessor, the dual resonance model. For a more detailed
presentation of these points see \cite{foun}. Every explicit computation of
the (graded) commutator of two string fields carried out by string theorists
has confirmed the infinite component pointlike nature \cite{Mar}\cite{Lowe},
but there is a strange ideological spirit which pervades the string community
which prevents them from saying clearly what they really compute. Reading the
two cited papers is a strange experience because it shows that correct
computations in times of a dominating metaphorical idea are no guaranty for a
correct interpretation. The authors come up with all kinds of metaphoric ideas
(including that of a string of which one only sees a point) in order to avoid
having to say "infinite component pointlike field" which would place them
outside their community.

Any philosophically motivated historian who wants to understand the Zeitgeist
which led to string theory and its various revolutions in the service of a
theory of everything, should find these (computationally correct but
conceptually strange) papers a rich source of information. Less than 7 decades
after Bohr and Heisenberg removed the metaphoric arguments of the old quantum
theory by introducing the concept of observables, the discourse within the
string theory community is trying to re-introduce metaphoric arguments into
the relativistic particle discourse. Surely one does not want to miss the kind
of fruitful transient metaphors which at the end led to valuable insights, but
what is a reasonable attitude with respect to an obviously incorrect metaphor
which hovers over particle theory ever since its beginnings in the 70s?

\section{String-localization and gauge theory}

Zero mass fields of finite helicity play a crucial role in gauge theory.
Whereas in classical gauge theory a pointlike massless vectorpotential is a
perfectly acceptable concept, the situation changes in QT as a consequence of
the Hilbert space positivity, which for massless unitary representations leads
to the loss of many spinorial realizations (as expressed in the second line of
(\ref{line})), in particular to that of the vector-potential without which it
is hardly possible to formulate perturbative QED. The traditional way to deal
with this situation has been to allow vector-potentials to live in an
indefinite metric space and to add ghost degrees of freedom in intermediate
calculations in such a way that the physical objects in form of the local
observables in a Hilbert space coalesce with the gauge or BRST invariant
objects under a suitably defined gauge or BRST group action. 

The ghost degrees of freedom are like catalyzers in chemistry; they are not
there in the initial set up of the problem and they have gone at the end, but
without there presence the renormalized perturbative pointlike field formalism
would not work for zero mass spin $s\geq1$ quantum matter. At the bottom of
the problem is a clash between modular localization and the Hilbert space
structure: although the ($m=0,s\geq1$) representations are pointlike generated
(\ref{line}) there are only field-strength but no potential-type generators
with $\left\vert A-\dot{B}\right\vert >s.$ Gauge theory tries to resolve this
clash by using a catalyzer which violates the Hilbert space setting of QT.

Despite the undeniable success of this kind of quantum adaptation of the
perturbative gauge setting, there are two arguments against considering the
present formulation as the end of the story. One is of a more philosophical
kind and the other points towards a serious limitation of the gauge formalism.
From a philosophical point of view this setting violates the maxim of Bohr and
Heisenberg that one should always look for a formulation in which the
computational steps (and not only the final result) can be formulated in terms
of observables. More tangible is the objection that the existing gauge
formalism aims only at \textit{local} observables. There are interacting
generators of physical objects which do not admit pointlike generators but
whose sharpest possible localization is semiinfinite stringlike; the most
prominent ones are electric charge-carrying operators \cite{Bu}. Their
construction is \textit{not} part of the standard perturbative formalism but
they have to be defined "by hand".

The best localization for a charged generating field is that of a semiinfinite
Dirac-Jordan-Mandelstam string characterized \textit{formally} by the
well-known expression%
\begin{equation}
\Psi(x,e)=~"\psi(x)e^{\int_{0}^{\infty}ie_{el}A^{\mu}(x+\lambda e)d\lambda
}"\label{DJM}%
\end{equation}
Using a version of perturbation theory which was especially designed to
incorporate this formal DJM expression into the n$^{th}$ order renormalization
setting, Steinmann \cite{Stein} succeeded to attribute a renormalized
perturbative meaning to this formal expression. Connected with this
nonlocality aspect is the subtle relation of electrically charged fields to
charged particles which shows up in infrared divergencies of on mass shell
objects. In addition a charged particle, even after a long time of having left
the scattering region, will never completely escape the region of influence of
infrared real (not virtual!) photons whose energy is below the (arbitrarily
small but nonvanishing) registering resolution and which therefore remain
"invisible" (in the sense of unregistered). This makes charge particles
"infraparticles" i.e. objects whose scattering theory does not lead to
scattering amplitudes but only to inclusive cross sections.

The infrared divergence problems in QED, first studied in a simpler model by
Bloch and Nordsiek, whose phenomenological remedy required to trade scattering
amplitudes with inclusive cross section \cite{YFS}, turned out to have a very
profound conceptual explanation: the Hilbert space of QED does not contain an
irreducible representation with a sharp mass, less so can it be written in
terms of antisymmetric tensor products of such states; rather the electron
two-point function starts with a cut at $m_{e}$ which depends on $e.$ For this
to occur the presence of zero mass particles is necessary but not sufficient.
Their coupling for low energies must also be sufficiently strong, a
requirement which is fulfilled for the minimal coupling of photons in QED but
not for renormalizable couplings of $(m=0,s=0,1/2)\ ($e.g. not for the $\pi
$-$N$ coupling with massless pions). Also the converse holds, if the theory
allows for one-particle states in the sense that the theory has a mass-shell
than even if this mass shell is not separated from the continuum by a gap) the
theory possess a standard (LSZ) scattering theory \cite{Dy}.

For global gauge symmetries, the idea that the local observables in their
vacuum representation determines all charged representation and, by suitably
combining them, lead to the physical charged fields, was one of the most
seminal conceptual conquests in local quantum physics \cite{Haag}. The
superselected charge-carrying fields are in this way (up to some conventions)
uniquely determined in terms of the vacuum representation of the local
observables. In $d\geq4$ these fields are Bose/Fermi fields which act
irreducibly in a Hilbert space which contains all superselected sectors
associated to the system of local observables. They transform according to a
compact internal symmetry group whose existence is preempted by the
inconspicuous presence of a copy of the dual of a group within the net of
local observables; the latter in turn is the is the fixpoint subalgebra of the
field algebra under the action of the internal symmetry (always global). 

Each compact group (with the exception of supersymmetry) can appear as the
internal symmetry of a QFT. With this structural insight a long path of the
somewhat mysterious\footnote{This concept, which is central to local quantum
physics, does not exist in classical physics; but as any quantum object it can
be "red back" therein (see also classical Grassmann variables from reading
back Fermions).} concept of internal symmetries, which begun with Heisenberg's
$SU(2)$ isospin in nuclear physics, came to a beautiful conclusion. Global
symmetry groups are a tool by which the quantum locality principle arranges
the various local superselection rules of a given observable algebra belonging
to different superselection charges into a \textit{field algebra}. What
started so mysteriously with Heisenberg's isospin in the 30s, ended
brilliantly with the DHR superselection theory and its local presentation in
terms of superselection charge-carrying local fields; showing again what the
quantum causal localization principle is capable to achieve but also how
conceptually demanding it is to find that path which reduces an observed
property to its conceptual roots.

In d=3,2, the commutation relations may be \textit{plektonic} or
\textit{solitonic}, meaning that the fields obey braid group or soliton
commutation relations which require semiinfinite stringlike localization and
lead to a generalized spin\&statistics theorem \cite{Mu1} and to a situation
in which the internal and spacetime symmetries allow no clearcut separation
among them. But as before, the net of local observables determines modulo some
conventions its full field algebra which incorporates the superselected
charge-carrying fields. In both cases, the low- and higher- dimensional case,
there is no better characterization for the inverse problem \textit{neutral}
\textit{observables }$\rightarrow$\textit{ charge-carrying fields} than the
metaphor which Mark Kac used in connection with an acoustic inverse problem:
"how to hear the shape of a drum".

It is a natural question to ask whether this reconstruction permits a more
concrete formulation in the form of reconstructing bilocals by breaking up
local expressions as the electric current $\bar{\psi}(x)\gamma_{\mu}\psi(x)$
in an analogous manner as was done in the 60s in order to reconstruct bilocals
$A(x)A(y)$ from Wick-ordered locals :$A^{2}(x):$ via a lightlike limiting
process \cite{La-S}. In the case that the local operators are associated with
a local gauge theory as QED, one expects bilocals with "gauge-bridges" between
the two points. The partial results on this problem are scarce but encouraging
\cite{Jacobs}. It would be a major progress in gauge theory if electrically
charged bilocals including gauge bridges could be obtained from local currents
by such a lightlike splitting, so that formally the stringlike DJM charge
generating field (\ref{DJM}) appears in the limit of dumping one charge at infinity.

The problem of possible presence of interacting nonlocal generating fields in
the physical Hilbert space becomes more serious in theories involving
vectorfields coupled among themselves. Whereas one believes to have a physical
understanding of the local (=$~$gauge invariant) composites (whose
perturbation expansion in terms of invariant correlation functions has
incurable infrared divergencies\footnote{Only if perturbation theory is
formulated in a pure algebraic setting and the problem of states is treated as
a second step there is a chance to control the infrared divergencies.}), there
is no convincing idea about the conceptual status of the degrees of freedom
which are the analogs of the charged fields in QED. For many decades we have
been exposed to such evocating metaphoric words as quark- and gluon-
confinement. Whereas such ideas are quite natural in QM where they point to
enclosing quantum matter in a potential vault, QFT has no mechanism of hiding
degrees of freedom by localizing them. The only mechanism through which
degrees of freedom may escape observations in a theory in which localization
is the dominating physical principle is a \textit{weakening of localization}
i.e. the opposite of a quantum mechanical vault\footnote{The use of lattice
theory has also its limitations; for example there is no lattice description
of infraparticles.}. The delocalization of electrically charged particles due
to surrounding photon clouds in QED is obviously not sufficient. What one
needs is the understanding of a situation in which the gluon plays a double
role: that of a charge carrier and that of the photons hovering around it.
Contrary to the formal DJM expression for charged fields, there is little
chance that the formal spacetime structure of such a situation can be guessed
by analogies.

After this long informative detour let use return to the localization-QT clash
one confronts in passing from classical vectorpotentials to their quantum
counterpart and ask whether instead of the indefinite metric "catalyzer" we
could have chosen one which sacrifies the pointlike localization (for which
there is no physical support anyhow in view of what has been said about
charged fields) and instead works with what we obtained from combining the
Wigner representation theory with modular localization (section 6,7)? 

If we succeed to set up a renormalized perturbation theory in terms of
string-localized potentials and explain the extreme delocalization of charges
in terms of string-localized vector potentials delocalizing massive complex
matter fields through their QED interaction in a persistent way, we would not
need to talk about formalism-preserving "catalyzers" to start with. Instead we
would have an alternative formulation in which the gauge invariant locals are
identical to the pointlike generated charge neutral subalgebra of an algebra
which contains all string-localized charged fields. 

So using string-localized potentials $A_{\mu}(x,e)$ from the start we would
understand that the ghosts in Gupta-Bleuler and BRST were the prize to pay for
insisting in setting up perturbation theory with only pointlike fields against
our better structural knowledge that charged objects are necessarily
noncompact localized and that point-localized vectorpotentials not only go
against the Hilbert space setting, but also leave the origin of the
string-localized quantum Maxwell charges in terms of properties of the
interaction density shrouded in mystery. It is important not to be
misunderstood on this point; we are not saying the the gauge theory setting is
incorrect, it remains correct for all pointlike gauge invariant fields which
are composites of field strength and charge neutral matter fields but it fails
for charge-carrying fields. 

One could think that one would also be able to compute nonlocal gauge
invariants, which in the setting of the BRST formalism would amount to
determine invariants under the nonlinear acting BRST transformation; but this
an impossible task. This should be no surprise since in this step physical
nonlocals have to "pop out" of unphysical "pointlike" generated objects. The
parenthesis indicate that pointlike is not meant in a physical but only a
formal sense. But for the pointlike generated subalgebra of charge neutrals
the gauge approach is efficient (perhaps apart from Yang-Mills theories) and
based on a well-studied renormalization formalism but, since different from
the classical interaction which is perfectly consistent with the principle of
classical field theory, the quantized version leaves the Hilbert space one
needs the quantum gauge formalism to recover it. 

All this can be avoided in a setting where one couples string-localized
potentials $A(x,e)$; there are no ghosts and therefore there is no need for a
gauge formalism, all objects have an intrinsic string localization and the
pointlike localized form a subset. Describing the so constructed theory in
terms of field strengths instead of potentials the only stringlike generators
are the charged fields. As was explained in the section on modular
localization (\ref{line}) one can find intertwiners from the Wigner
representation to all spinor representations if one admits string-localized
potentials. In particular the one with the string-localized field of lowest
Lorentz spin ($A_{\mu}(x,e)$ for helicity h=1, $g_{\mu\nu}(x,e)$ for h=2,..)
has the lowest short distance dimension namely sdd=1 and hence the optimal
behavior from the viewpoint of renormalization theory. 

It is quite surprising that there are string-localized potentials  for
arbitrary spin s with scale dimension=1 which is the power counting
prerequisite for encountering renormalizable interactions. So by avoiding to
impose the unphysical restriction to pointlike interactions one also enlarges
the scope of renormalizability. A formalism with this enormous range cannot be
expected cannot be expected to fall into one's lap, in fact it is presently
still in its infancy, it poses completely new and largely unsolved problems.
But before commenting on this new task, it is helpful to delineate what one
expects of such an alternative approach.

Superficially the use of such string-localized fields seem to be
indistinguishable from the axial gauge\footnote{The axial gauge is the only
one which (after adjusting the Lorentz-transformation property of $e$) is
covariant and Hilbert-space compatible. So it never was a gauge in the sense
of sacrificing the Hilbert space structure. A renormalized perturbation theory
was not possible because its serious infrared problems were not understood as
indicating string- instead of point-localization.}; in both cases the
conditions $\partial^{\mu}A_{\mu}(x,e)=0=e^{\mu}A_{\mu}(x,e)$ are obeyed. In
the axial gauge interpretation the $e$ is a gauge parameter and does not
participate in Lorentz transformations, whereas in case of string-localized
field the spacelike unit vector transforms as a string direction or, what is
the same, as a point in a 3-dimensional de Sitter spacetime. The axial gauge
failed as a perturbative computational tool in the standard setting as a
result of its incurable infrared divergence problems. In a way the
string-localized approach explains this as a consequence of quantum
fluctuations \textit{both} in $x$ and $e$ which makes it necessary to use
testfunction smearing in $x$ and $e$ and to discuss coalescing point limits
with the same care as for composite fields. The guiding idea is that the use
of string dependent potentials delocalizes the charged field automatically so
that there is no necessity to use ad hoc formulas as (\ref{DJM}) and to engage
in the difficult task to construct their renormalized counterpart. 

Whereas the standard renormalization formalism for pointlike fields admits
many different variations of which the Epstein Glaser scheme is the one which
uses causal locality most heavily\footnote{The other formulations based on
Lagrangians and Euclidean continuations use the close relations between the
classical fields and their Euclidean counterparts.} But for the
string-localized approach which admits no Lagrangian formulation, the
Epstein-Glaser \cite{E-G} is the only one. In the pointlike case the knowledge
of the $n^{th}$ order determines the $n+1$ order up to a term on the total
diagonal which limits the freedom to the addition of pointlike composites. 

If the power counting requirement for renormalizability is not only necessary
but also sufficient one would have an enormous enrichment of renormalizable
interactions generalizing gauge theories or rather its string-localized
reformulation. So this would open a vast new area of research. The interesting
question is then whether not only interations of zero mass higher spin
particles with low spin massive matter but also massless higher spin particles
can interact among themselves. For s=1 the only known models are the
Yang-Mills theories; one expects for s=2 models whose Riemann tensor-like
pointlike field strengths are nonlinear expressions in string-like $g_{\mu\nu
}(x,e)$ tensor potentials. In both cases the existence of pointlike composites
of stringlike fields is the guiding principle and not the group theoretical
structure of the interaction in terms of stringlike potentials. 

Up to now the issue was how to couple massless higher spin $s\geq1$ to low
spin massive matter. For massive higher spin there is no representation
theoretical argument to introduce string-localized potentials since all the
covariant possibilities are (\ref{line}) are realized. There is however the
power-counting requirement which goes against the use of the $A_{\mu}(x)$ with
$\partial^{\mu}A_{\mu}(x)=0$  since its short distance dimension is 2 instead
of one as a result of the additional degree of freedom which distinguishes the
massive case from its massless counterpart. Hence the pointlike vectorfield
has the same short distance dimension as its "field strength" and therefore
falls outside the power counting limit. There exists however a massive
string-localized potential $A_{\mu}^{(m)}(x,e)$ of $sddim.=1$ associated with
this field strength which for $m\rightarrow0$ passes to the zero mass
potential. Its use certainly complies with the power counting requirement, but
does its use lead to an acceptable physical theory?

By acceptable we mean a theory in which the zero order string localization
does not spread "all over the place" i.e. in which there exist still pointlike
generated subalgebras as those generated by charge-neutral fields in QED.
Fortunately there is an additional mechanism, which according to our present
knowledge seems to secure precisely this picture of the existence of local
observables: The Schwinger-Higgs screening mechanism. It is the QFT analog of
the Debeye screening in QM. The latter describes the transition from a long
range Coulomb system to a one in which the effective action falls off like a
Yukawa potential (the compensating effects of opposite charges. In QFT
screening would be a much more violent mechanism because because the analog of
range of forces is the spacetime localization of generating fields\footnote{A
mathematical theorem which explains the connection between the gain of
analyticity in 3-point functions indicating a gain in localization (screening)
and the conversion of photons into massive vectormesons can be found in
\cite{Swieca}. Swieca used the screening terminology in most of the
publications but it seems that this got lost during the 70s. }. 

It was Schwinger's idea that something like this could occur in actual
(spinor) QED and render the whole theory massive, but since he could not find
a perturbative argument, he invented the Schwinger model (massless QED in
d=1+1) which only leaves the screened phase and approaches the charged Jordan
model in the short distance limit \cite{Jor}. The contribution of Higgs
consists in a more interesting model which allows a perturbative version of
screening; the Higgs model in its original form is nothing but screened scalar
QED. The screening mechanism formulated on scalar QED, using the
string-localized vectorpotential maps this model to one with a massive
stringlike vectorpotential and a real scalar field, so that half the degrees
of freedom of the charged complex field was used to convert the massless
photon into a massive vectormeson. This process has nothing to do with what in
the literature is called spontaneous symmetry breaking (Goldstone). Certainly
screening leads to a symmetry reduction since the Maxwell charge is zero and
hence its superselection rule has been lost. The nonvanishing local
expectation of $\Phi$ is part of the prescription and of no intrinsic
significance; the intrinsic meaning is related to the conserved currents: the
divergence of the charge in the case of spontaneous symmetry-breaking (as the
result of the existence of a Goldstone Boson) and the vanishing of the charge
in the case of screening.  A more detailed description can be found in
\cite{char}.

The important question which remained unanswered in the 70s is whether this
screening mechanism is a peculiar illustration for how an interacting massive
vectormeson can be part of a pointlike local QFT or whether  this is a special
case of a more general intrinsic mechanism which states that in order to
maintain locality interacting massive higher spin particles must be
accompanied by lower spin objects? Different spins have been linked together
by the invention of supersymmetry, but it would be more natural to understand
this as a consequence of the locality principle. An supporting argument was
given within the BRST setting \cite{DS}: if one starts with a massive
vectormeson, the Higgs meson (but now with vanishing vacuum expectation) has
to be introduced for maintaining consistency of the BRST formalism. Only by
removing the non-intrinsic BRST formalism and using instead the stringlike
sdd=1 vector-potentials one can hope to understand the crucial role of
locality in a conjectured \textit{lower spin companion} mechanism behind the
Higgs issue.

\section{Building LQP via \textit{positioning of monads} in a Hilbert space}

We have seen that modular localization of states and algebras is an intrinsic
i.e. field-coordinatization-independent way to formulate the kind of
localization which is characteristic for QFT. It is deeply satisfying that it
also leads to a new constructive view of QFT.

\textbf{Definition }(Wiesbrock \cite{Wies1})\textbf{:} \textit{An inclusion of
standard operator algebras }$\left(  \mathcal{A\subset B},\Omega\right)
$\textit{ is "modular" if }$\left(  \mathcal{A},\Omega\right)  $\textit{ and
(}$\mathcal{B}$\textit{,}$\Omega$\textit{) are standard and }$\Delta
_{\mathcal{B}}^{it}$\textit{ acts like a compression on }$\mathcal{A}$\textit{
i.e. }$Ad\Delta_{\mathcal{B}}^{it}\mathcal{A}\subset\mathcal{A}.$\textit{ A
modular inclusion is said to be standard if in addition the relative commutant
(}$\mathcal{A}^{\prime}\cap\mathcal{B},\Omega$\textit{) is standard. If this
holds for }$t<0$\textit{ one speaks about a -modular inclusion.}

The study of inclusions of operator algebras has been an area of considerable
mathematical interest. Particle physics uses three different kind of
inclusions; besides the modular inclusions, which play the principal role in
this section, there are \textit{split inclusions} and inclusions with
conditional expectations (or using the name of their protagonist, Vaughn
\textit{Jones inclusions).} Split inclusions play an important role in
structural investigation and are indispensable in the study of thermal aspects
of localization, notably localization entropy (see second part). Inclusions
with conditional expectations result from reformulating the DHR theory of
superselection sectors which in its original formulation uses the setting of
localized endomorphisms of observable algebras \cite{Haag}.

Inclusions $\mathcal{A}\subset\mathcal{B}$ with conditional expectation
$E(\mathcal{B})$ cannot be modular and the precise understanding of the reason
discloses interesting insights. According to a theorem of Takesaki \cite{Tak}
the existence of a conditional expectation is tantamount to the modular group
of the smaller algebra being equal to the restriction of that of the bigger.
Hence the natural generalization of this situation is that the group
$Ad\Delta_{\mathcal{B}}^{it}$ of the larger algebra acts on $\mathcal{A}$ for
either $t<0$ or for $t>0$ as a bona fide compression (endomorphism) which
precludes the existence of a conditional expectation. Intuitively speaking
modular inclusions are "too deep" to allow conditional expectations.
Continuing this line of speculative reasoning one would expects that inasmuch
as "flat" inclusions with conditional expectations are related to inner
symmetries, "deep" inclusions of the modular kind should lead to spacetime symmetries.

This rough guess turns out to be correct. The main aim of modular inclusions
is really twofold, on the one hand to \textit{generate spacetime symmetry}
which than acts on the original algebras and creates a \textit{net of
spacetime indexed algebras} which are covariant under these symmetries. For
the above modular inclusion of two algebras this is done as follows: from the
two modular groups $\Delta_{\mathcal{B}}^{it},\Delta_{\mathcal{A}}^{it}$ one
can form a unitary group $U(a)$ which together with the modular unitary group
of the smaller algebra $\Delta_{\mathcal{B}}^{it}$ leads to\ the commutation
relation $\Delta_{\mathcal{B}}^{it}U(a)=U(e^{-2\pi t}a)\Delta_{\mathcal{B}%
}^{it}$ which characterizes the 2-parametric translation-dilation (Anosov)
group. One also obtains a system of local algebras by applying these
symmetries\ to the relative commutant $\mathcal{A}^{\prime}\cap\mathcal{B}.$
From these relative commutants one may form a new algebra $\mathcal{C}$%
\begin{equation}
\mathcal{C}\equiv\overline{\bigcup_{t}Ad\Delta_{\mathcal{B}}^{it}%
(\mathcal{A}^{\prime}\cap\mathcal{B})}%
\end{equation}
In general $\mathcal{C}\subset\mathcal{B}$ and we are in a situation of a
nontrivial inclusion to which the Takesaki theorem is applicable (the modular
group of $\mathcal{C}$ is the restriction of that of $\mathcal{B})$ which
leads to a conditional expectation $E:\mathcal{B}\rightarrow\mathcal{C}$;
$\mathcal{C}$ may\ also be trivial. The most interesting situation arises if
the modular inclusion is \textit{standard }i.e. all three algebras
$\mathcal{A},\mathcal{B},\mathcal{A}^{\prime}\cap\mathcal{B}$ are standard
with respect to $\Omega;$ in that case we arrive at a chiral QFT.

\textbf{Theorem}: (Guido,Longo and Wiesbrock \cite{G-L-W}) \textit{Standard
modular inclusions are in one-to-one correspondence with strongly additive
chiral LQP.}

Here chiral LQP is a net of local algebras indexed by the intervals on a line
with a Moebius-invariant vacuum vector and \textit{strongly additive} refers
to the fact that the removal of a point from an interval does not
\textquotedblleft damage\textquotedblright\ the algebra i.e. the von Neumann
algebra generated by the two pieces is still the original algebra. One can
show via a dualization process that there is a unique association of a chiral
net on $S^{1}=\mathbb{\dot{R}}$ to a strongly additive net on $\mathbb{R}$.
Although in our definition of modular inclusion we have not said anything
about the nature of the von Neumann algebras, it turns out that the very
requirement of the inclusion being modular forces both algebras to be
hyperfinite type III$_{1}$ algebras.

The closeness to Leibniz's idea about (physical) reality of originating from
relations between monads (with each monad in isolation being void of
individual attributes) more than justifies our choice of name; besides that
"monad" is much shorter than the somewhat long winded mathematical terminology
"hyperfinite type III$_{1}$ Murray-von Neumann factor algebra". The nice
aspect of chiral models is that one can pass between the operator algebra
formulation and the construction with pointlike fields without having to make
additional technical assumptions\footnote{The group theoretic arguments which
go into that theorem \cite{Joerss} seem to be available also for higher
dimensional conformal QFT.}. Another interesting constructive aspect is that
the operator-algebraic setting permits to establish the existence of algebraic
nets in the sense of LQP for all $c<1$ representations of the energy-momentum
tensor algebra. This is much more than the vertex algebra approach is able to
do since that formal power series approach is blind against the dense domains
which change with the localization regions.

The idea of placing the monad into modular positions within a common Hilbert
space may be generalized to more than two copies. For this purpose it is
convenient to define the concept of a \textit{modular intersection} in terms
of modular inclusion.

\textbf{Definition (}Wiesbrock \cite{Wies1}\textbf{)}: \textit{Consider two
monads }$A$\textit{ and }$B$\textit{ positioned in such a way that their
intersection }$A\cap B$\textit{ together with A and B are in standard position
with respect to the vector }$\Omega\in H$\textit{. Assume furthermore}%

\begin{align}
&  (\mathcal{A}\cap\mathcal{B\subset}\mathcal{A)~}and~(\mathcal{A}%
\cap\mathcal{B\subset B)~}are~\pm mi\\
&  J_{\mathcal{A}}\lim_{t\rightarrow\mp}\Delta_{\mathcal{A}}^{it}%
\Delta_{\mathcal{B}}^{-it}J_{\mathcal{A}}=\lim_{t\rightarrow\mp}%
\Delta_{\mathcal{B}}^{it}\Delta_{\mathcal{A}}^{-it}\nonumber
\end{align}
\textit{then (}$A,B,\Omega$\textit{) is said to have the }$\pm$\textit{
modular intersection property (}$\pm~$\textit{mi)}.

It can be shown that this property is stable under taking commutants i.e. if
$\left(  \mathcal{A},\mathcal{B},\Omega\right)  \pm mi$ then $\left(
\mathcal{A}^{\prime},\mathcal{B}^{\prime},\Omega\right)  $ is $\mp mi.$

The minimal number of monads needed to characterize a 2+1 dimensional QFT
through their modular positioning in a joint Hilbert space is three. The
relevant theorem is as follows

\textbf{Theorem}: (Wiesbrock \cite{Wies2}) \textit{Let }$A_{12}$%
\textit{,}$A_{13}$\textit{ and }$A_{23}$\textit{ be three monads\footnote{As
in the case of a modular inclusion, the monad property is a consequence of the
modular setting. But for the presentation it is more convenient and elegant to
talk about monads from the start.} which have the standardness property with
respect to }$\Omega\in H$\textit{. Assume furthermore that}%
\begin{align}
&  (\mathcal{A}_{12},\mathcal{A}_{13},\Omega)~is~-mi\\
&  (\mathcal{A}_{23},\mathcal{A}_{13},\Omega)~is~+mi\nonumber\\
&  (\mathcal{A}_{23},\mathcal{A}_{12}^{\prime},\Omega)~is~-mi\nonumber
\end{align}

\textit{then the modular groups }$\Delta_{12}^{it}$\textit{, }$\Delta
_{13}^{it}$\textit{ and }$\Delta_{23}^{it}$\textit{ generate the Lorentz group
}$SO(2,1)$\textit{.}

Extending this setting by placing an additional monad $\mathcal{B}$ into a
suitable position with respect to the $\mathcal{A}_{ik}$ of the theorem, one
arrives at the Poincar\'{e} group $\mathcal{P}(2,1)$ \cite{Wies3}$.$ The
action of this Poincar\'{e} group on the four monads generates a spacetime
indexed net i.e. a LQP model and all LQP have a monad presentation.

To arrive at d=3+1 LQP one needs 6 monads \cite{K-W}. The number of monads
increases with the spacetime dimensions. Whereas in low spacetime dimensions
the algebraic positioning is natural within the logic of modular inclusions,
in higher dimensions it is presently necessary to take some additional
guidance from geometry, since the number of possible modular arrangements for
more than 3 monads increases. There is an approach with similar aims of
characterizing a QFT by its modular data by Buchholz and Summers \cite{CGMA}.
Instead of the modular groups these authors use the modular reflections $J.$
For our purpose of characterizing local quantum physics in terms of
positioning of monads the approach proposed by Wiesbrock based on modular
inclusions and intersections is more convenient. Its orgin dates back to the
observation that the Moebius group can be extracted from the modular groups of
the quarter circle algebras \cite{circle}.

We have presented these mathematical results and used a terminology in such a
way that the relation to Leibniz philosophical view about relational reality
is visible.

This is not the place to give a comprehensive account, but only an attempt to
direct the attention of the reader to this (in my view) startling conceptual
development in the heart of QFT, a theory which despite its almost 90 years of
existence is still far from its closure..

Besides the radically different conceptual-philosophical outlook on what
constitutes QFT, the modular setting offers new methods of construction. For
that purpose it is however more convenient to start from one monad
$\mathcal{A}\subset B(H)$ and assume that one knows the action of the
Poincar\'{e} group via unitaries $U(a,\Lambda)$ on $\mathcal{A}.$ If one
interprets the monad $\mathcal{A}$ as a wedge algebra $\mathcal{A=A(}%
W\mathcal{)}$ than the Poincar\'{e} action generates a net of wedge algebras
$\left\{  \mathcal{A(}W\mathcal{)}\right\}  _{W\in\mathcal{W}}.$ A QFT is
supposed to have local observables and hence if the double cone
intersections\footnote{Double cones are the typical causally complete compact
regions which can be obtained by intersecting wedges.} $\mathcal{A(}%
\mathcal{D}\mathcal{)}$ turn out to be trivial (multiples of the identity
algebra), the net of wedge algebras does not leads to a QFT. This is expected
to be the algebraic counterpart of a Lagrangian which does not have a have a
corresponding QFT. If however these intersections are nontrivial, we would
have a rigorous existence proof; the existence of a generating field for those
double cone algebras is then merely a technical problem. There are of course
two obvious sticking points: (1) to find the action of the Poincar\'{e}
on\ $\mathcal{A(}W_{0}\mathcal{)}$\ and (2) a method which establishes the
non-triviality of intersections of wedge algebras and leads to formulas for
their generating elements.

As was explained in the previous section, both problems have been solved
within a class of factorizing models \cite{Lech1}. Nothing is known about how
to address these two points in the more general setting i.e. when the tempered
PFG are not available.

The monad setting has only been formulated for Poincar\'{e}-covariant QFT. A
extension to locally covariant QFT in CST is expected to present a new path
towards the still elusive Quantum Gravity. It is tempting to think of the
diffeomorphisms of AQFT in CST to be of modular origin. A particularly simple
illustration is $\ Diff(S^{1}),$ the diffeomorphism group of chiral theories
on a circle. It is well known that the vacuum is only invariant under the
Moebius subgroup and there are no states which are invariant under higher
diffeomorphisms. The candidates for the higher modular groups are the
diffeomorphisms which fix more than two points which can be obtained from a
covering construction involving roots of fractional M\"{o}bius
transformations. The resulting multi-interval construction suggests to look
for the modular group of a multi-interval; the problem is to find the
appropriate states which lead to a geometric modular group. This problem was
solved very recently by Longo, Kawahigashi and Rehren \cite{LKR}. The
interesting aspect of their solution (in agreement with the absence of
eigenstates of higher diffeomorphisms) is that the resulting modular groups
are only partially geometric i.e. geometric only inside the multi-interval.
This is of course what one expects in the case of isometries in CST.

Another interesting problem which is on the verge of being solved is the
existence of a higher (m=0,s%
$>$%
1)  quantum Aharonov-Bohm effect. The quantum A-B effect in the setting of
AQFT is the statement that the  electromagnetic free quantum field shows a
violation of Haag duality \cite{LRT} for a non simply connected toroidal
spacetime region $\mathcal{T}$%
\begin{equation}
\mathcal{A}(\mathcal{T})\subset\mathcal{A}(\mathcal{T}^{\prime})^{\prime}%
\end{equation}
whereas for simply connected regions the equality (Haag duality) holds. For
higher spin massive fields Haag duality holds for any region. The A-B
interpretation is that that the right hand side contains observables which
cannot be constructed from field strengths in the torus. This violation of
Haag duality has been shown in an old unpublished work before the modular
methods became available. A modular approach to this problem yields more than
just the violation of Haag duality, one also can compute a modular group and
there is a close relation to the previous 4-fix point problem. What makes this
problem so fascinating is the fact that it has a nontrivial extension to zero
mass s%
$>$%
1 in which case higher genus A-B fluxes result. So it places s=1 gauge theory
and the higher spin extensions on the same A-B footing.

Finally we should mention one unsolved long-lasting issue of modular theory:
the modular group of the free massive double cone algebra (with respect to the
vacuum) which is known to act "fuzzy" (non-geometric) and has been conjectured
to have a Hamiltonian which acts as a pseudo-differential operator instead of
a differential operator \cite{S-W}. There are rather convincing arguments that
the holographic projection of such a situation leads to a geometric modular
movement on the horizon \cite{Hol}. This suggested the idea that if one knew a
formula for the propagation of characteristic massive data on the horizon into
the inside of the double cone, the fuzzy action may simply come about by
applying this formula to the geometric group on the horizon. Such a formula
has recently appeared in \cite{Bis}%
\begin{equation}
A_{m}(x)=-2i\int_{LF}dy_{+}d^{2}y_{\perp}\Delta_{m}(x-y)|_{y_{-}=0}%
A(y_{+},y_{\perp})
\end{equation}
where $A(y_{+},y_{\perp})$ is the transverse extended chiral holographic
projection of the massive bulk field $A_{m}(x).$

A scale tranformation on $A(y_{+},y_{\perp})$ acts on $y_{+}$ and since
$y_{-}=0$ we can apply the inverse scale transformation to $y_{-}$ without
changing anything. By renaming variables we can maintain the original unscaled
variable in $A$ if we replace the y in $\Delta_{m}$ by $y_{+}\rightarrow
\lambda^{-1}y_{+},y_{-}\rightarrow\lambda y_{-}.~$\ Using the Lorentz
invariance of $\Delta_{m}$ we may shift this transformation to the $x$. So the
upshot is that the dilation on $H(W)$ passes to the Lorentz boost on the bulk
$W.$

Let us now see how the "fuzzyness" develops in the case of a double cone. For
simplicity we stay in $d=1+1$ and chose a double cone symmetric around the
origin as in \cite{Haag}. Then the lower mantle of the cone with apex $(-1,0)$
is a Horizon whose causal shadow covers the double cone. Every signal which
entered the double cone must have entered through the mantle. In this case the
propagation from the two pieces of the mantle leads to the sum%
\begin{align}
&  A_{m}(x_{+},x_{-})=-2i\int_{-1}^{+1}dy_{+}\Delta_{m}(x-y)|_{y_{-}=0}%
A(y_{+})+\\
&  +-2i\int_{LF}dy_{-}\Delta_{m}(x-y)|_{y_{+}=0}A(y_{-})\nonumber
\end{align}
Now the modular group on the Horizon is fractional namely the "dilation" which
leaves the fixed points $y_{\pm}=-1,+1$ invariant instead of $0,\infty$ as in
the first case. The modular group on both parts of the horizon is%
\begin{equation}
x_{\pm}(s)=\frac{(1+x_{\pm})-e^{-s}(1-x_{\pm})}{(1+x_{\pm})+e^{-s}(1-x_{\pm})}%
\end{equation}
Different from the previous case one cannot transfer this fractional change
from the y to the x. This time there is no local transformation, rather the
action on $A_{m}(x)$ is fuzzy but stays inside the double cone. It is however
not purely algebraic since it was obtained by combining the geometric group on
the Horizon with the causal propagation whose reverberation aspect causes the
fuzzyness. It can be shown that as in the case of a wedge the interaction does
not change anything, it is always this semi-geometric modular group. A more
general discussion including a calculation of the pseudodifferential
generators of these modular actions will be contained in a forthcoming paper
by Brunetti and Moretti.

So it looks that there is some new movement on this long-lasting issue.

\section{}

\end{document}